\documentclass{elsart}

\usepackage{amssymb}
\usepackage{graphicx}
\usepackage{mathrsfs}

\def\frak#1#2{\textstyle\frac{#1}{#2}}
\def\half{\textstyle\frac{1}{2}}
\newcount\Figsinline \def\Figsinline{1}

\newcommand{\Figure}[4]
{ \begin{figure}
  \begin{center}
     \includegraphics[scale=#2]{#1} 
  \end{center}
  \caption{\small #4} \label{#3}
  \end{figure} }

\def\sn{\mathrm{sn}}
\def\cn{\mathrm{cn}}
\def\dn{\mathrm{dn}}

\begin{document}

\begin{frontmatter}



\title{Pulsation and Precession of the Resonant Swinging Spring}


\author{Peter Lynch\corauthref{cor1}}
\ead{Peter.Lynch@met.ie}
\ead[url]{http://www.maths.tcd.ie/$\sim$plynch}
\corauth[cor1]{Corresponding author} 
\address{Met \'Eireann, Glasnevin Hill, Dublin 9, Ireland}

\author{Conor Houghton}
\ead{houghton@maths.tcd.ie}
\ead[url]{http://www.maths.tcd.ie/$\sim$houghton}
\address{Department of Mathematics, Trinity College, Dublin 2, Ireland}

\begin{abstract}

When the frequencies of the elastic and pendular oscillations
of an elastic pendulum or swinging spring are in the ratio two-to-one,
there is a regular exchange of energy between the two modes of
oscillation. We refer to this phenomenon as {\it pulsation}.
Between the horizontal excursions, or pulses,
the spring undergoes a change of azimuth which we call the 
precession angle.  The pulsation and stepwise precession are the
characteristic features of the dynamics of the swinging spring.

The modulation equations for the small-amplitude resonant 
motion of the system are the well-known three-wave equations.
We use Hamiltonian reduction to determine a complete analytical 
solution. The amplitudes and phases are expressed in terms of both
Weierstrass and Jacobi elliptic functions. The strength of
the pulsation may be computed from the invariants of the equations.
Several analytical formulas are found for the precession angle.

We deduce simplified approximate expressions, in terms of elementary
functions, for the pulsation amplitude and precession angle
and demonstrate their high accuracy by numerical experiments. 
Thus, for given initial conditions, we can describe the envelope
dynamics without solving the equations.
Conversely, given the parameters which determine the envelope,
we can specify initial conditions which, to a high level of
accuracy, yield this envelope.

\end{abstract}

\begin{keyword}
elastic pendulum \sep swinging spring \sep nonlinear resonance \sep
three-wave equations \sep precession \sep pulsation\sep monodromy

\PACS 05.45.-a \sep 02.30.Ik \sep 45.20.Jj
\end{keyword}
\end{frontmatter}


\section{Introduction}
\label{Sec:Introduction}

The present work is concerned with the three-dimensional motion 
of the elastic pendulum or swinging spring in the case of resonance.
It continues the investigation described in previous studies
by Lynch \cite{Lynch2002a} and by Holm and Lynch \cite{H&L}.
In particular, the
exchange of energy between quasi-vertical and quasi-horizontal
oscillations and the stepwise precession of the swing plane are
investigated.

When the ratio of the normal mode frequencies of the spring is 2:1,
a resonance occurs, in which energy is
transferred periodically between vertical and horizontal oscillations.
The first study of this resonance was that of Vitt and Gorelik \cite{V&G}.
We refer to the regular exchange phenomenon as {\it pulsation}.
The motion has two distinct characteristic times, 
that of the fast oscillations and that of the slow pulsation envelope.  
As the oscillations change from horizontal to vertical and back again,
it is observed that each horizontal excursion or pulse is in a 
different direction.
We call this change in azimuth the precession angle.
The motion thus has three components: oscillation (fast),
pulsation (slow) and precession (slow), closely analogous to
the rotation (fast), nutation (slow) and precession (slow) of a
spinning top \cite{Audin}.%
\footnote{A Java Applet illustrating the pulsation of the
swinging spring may be found at \hfil\break
{\tt
http://www.maths.tcd.ie/$\sim$plynch/SwingingSpring/SS\_Home\_Page.html}.}

\newcounter{listnum}
We consider two complementary questions, one direct and one inverse:
\begin{list}%
{{\it Question}~\arabic{listnum}.}%
{\usecounter{listnum}\setlength{\rightmargin}{\leftmargin}}
\item
Given initial conditions, can we describe the envelope
dynamics without solving the equations?
\item
Given the parameters which determine the envelope,
can we specify initial conditions which yield this envelope?
\end{list}%
We provide a complete answer to Question~1.  Analytical expressions
are derived for the pulsation amplitude, precession angle
and period in terms of the invariants of the motion.
We also develop accurate approximate expressions
for the pulsation amplitude and precession angle. Thus, the 
envelope dynamics may be deduced from the initial conditions.
Question~2 is more recondite, but we can give a positive answer for
the physically interesting case of strong pulsation.
We derive approximate expressions for the angular momentum
and Hamiltonian in terms of the pulsation amplitude and precession angle. 
Initial conditions can then be determined which yield the desired
envelope to a good level of approximation.

We briefly outline the contents of the paper below.
When the amplitude is small, the Lagrangian may be
approximated to cubic order. When it is averaged over the fast
oscillation time, a set of equations for the envelope amplitudes
is obtained. These modulation equations, the three-wave equations,
are presented in \S2. They are found to have three independent
constants of motion and are therefore completely integrable.
Small-amplitude perturbations about steady-state solutions are
studied in \S3, and a crude estimate of the precession angle
is obtained.  The general solution of the three-wave equations for
finite-amplitude motions is derived in \S4.
The amplitudes are expressed in terms of
elliptic functions and the phase angles as elliptic integrals.
Analytical expressions for the stepwise precession of
the swing-plane are then derived. 

Recently, Dullin {\it et al.} \cite{Dullin} constructed a canonical
transformation in which the angle of the swing plane is a coordinate
in an action-angle system. They showed that the precession angle is 
one of the two rotation numbers of the invariant tori of the integrable
system. They obtained a simple equation for the precession angle 
by approximating an elliptic integral.  They proved analytically that
the resonant swinging spring has monodromy and concluded that the
system provides a clear physical demonstration of this phenomenon.

Several approximate expressions for the precession angle, involving only
elementary functions, are obtained in \S5. One of these
is equivalent to the formula reported in \cite{Dullin}.
The approximate solutions are compared to the values obtained from the
analytical expression, and are found to give remarkably accurate results.
The intensity of the pulsation envelope is determined by 
solving a cubic equation whose coefficients are defined by the invariants.
Thus, the direct question is fully answered, in the affirmative.

To answer the inverse question, we assume the pulsation amplitude and
precession angle are given and derive expressions for the invariants. 
From these, appropriate initial conditions are easily determined.
The expression for the precession angle is easily inverted. To obtain an
invertible expression for the pulsation amplitude, we approximate 
the cubic by a quadratic, and obtain in \S6 simple approximate expressions
for the angular momentum and Hamiltonian.  These approximations may
be used to control the envelope dynamics by an appropriate choice of
initial conditions.

In the concluding section, \S7, we present a schematic diagram which
shows the qualitative dependence of the envelope motion on the values
of the invariants.  This allows us to determine, at a glance, the
general character of the solution for given values of the constants of
motion. Several important special solutions are indicated on the diagram.


\section{The Dynamical Equations}
\label{Sec:Dynamics}

%
\Figure{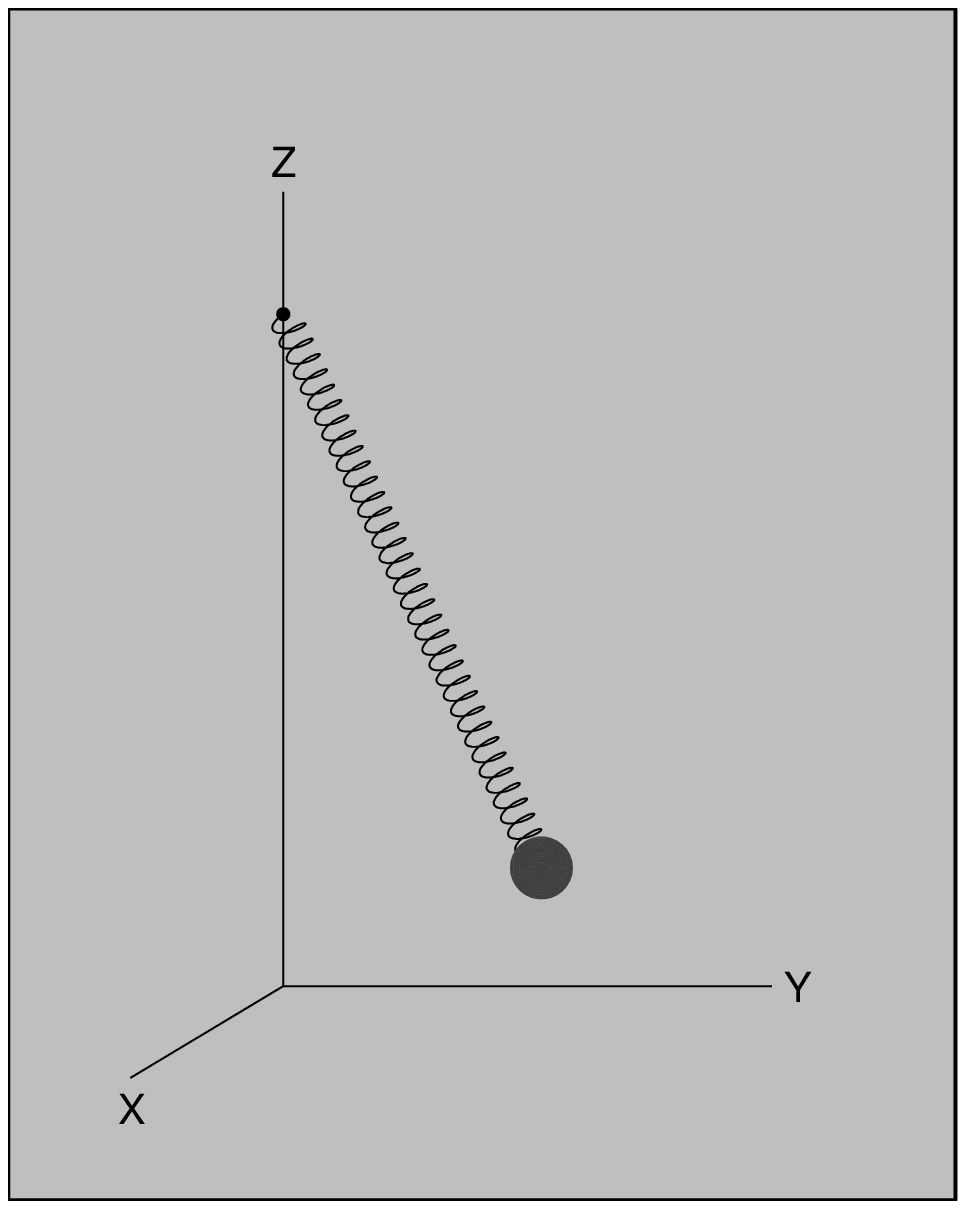}{0.70}{Fig:SwingingSpring}
{The swinging spring. Cartesian coordinates are
used, with the origin at the point of stable equilibrium of the bob.
The pivot is at point $(0,0,\ell)$.}
%

The physical system under investigation is an elastic pendulum, or
swinging spring, consisting of a heavy mass suspended from a fixed
point by a light spring and moving under gravity, $g$
(Fig.~\ref{Fig:SwingingSpring}).
We assume an unstretched length $\ell_0$, length $\ell$ at
equilibrium, spring constant $k$ and mass $m$.
The Lagrangian, approximated to cubic order in the
amplitudes, is
\begin{equation}\label{Lag-3D}
L = \half \left( \dot x^2+\dot y^2+\dot z^2 \right)
 -  \half \left[ \omega_R^2(x^2+y^2)+\omega_Z^2 z^2 \right]
 +  \half \lambda(x^2+y^2) z \, ,
\end{equation}
where $x$, $y$ and $z$ are Cartesian coordinates centered at the point
of equilibrium, $\omega_R=\sqrt{g/\ell}$ is the frequency of linear
pendular motion, $\omega_Z=\sqrt{k/m}$ is the frequency of its elastic
oscillations and $\lambda=\ell_0\omega_Z^2/\ell^2$.
The equations of motion in cartesian, spherical and cylindrical
coordinates may be found in \cite{Lynch2002a}. 
There are two constants of the motion, the total energy 
and the angular momentum about the vertical, and the system is not integrable.
Its chaotic motions have been studied by many authors (see
Refs. in \cite{Lynch2002b}).

\subsection{The Time-averaged Equations}

We confine attention to the resonant case $\omega_Z=2\omega_R$
and apply the averaged Lagrangian technique.
The solution is assumed to be of the form
\begin{eqnarray}
x &=& \Re\{a(t)\exp(i\omega_R t)\} \, ,  \label{ModSolA} \\
y &=& \Re\{b(t)\exp(i\omega_R t)\} \, ,  \label{ModSolB} \\
z &=& \Re\{c(t)\exp(2i\omega_R t)\} \,.  \label{ModSolC}
\end{eqnarray}
The coefficients $a(t)$, $b(t)$ and $c(t)$ are assumed
to vary on a time scale which is much longer than the time-scale of the
oscillations, $\tau = 1/\omega_R$.
The Lagrangian is averaged over this time, yielding
\[
\langle L \rangle = \half\omega_R
\Big[  \Im\{\dot a a^* + \dot b b^* + 2\dot c c^*\}
 + \Re\{\kappa(a^2+b^2)c^*\} \Big]
\,,
\]
where $\kappa=\lambda/(4\omega_{\rm R})$.
The resulting Euler-Lagrange equations are the modulation equations for
the envelope dynamics:
\begin{eqnarray}
i\dot a &=& \kappa a^*c  \,,      \label{niceone} \\
i\dot b &=& \kappa b^*c   \,,     \label{nicetwo} \\
i\dot c &=& \textstyle{\frac{1}{4}}\kappa (a^2+b^2) \,.
\label{nicethree}
\end{eqnarray}

\subsection{The three-wave equations}

We now transform to new variables
\begin{equation}
A = \half\kappa(a+ib)\,, \quad B = \half\kappa(a-ib)\,, \quad C = \kappa c
\,.
\end{equation}
Then the equations for the envelope dynamics take the form
\begin{eqnarray}
i\dot A &=& B^*C \, ,  \label{TWEa}   \\
i\dot B &=& CA^* \, ,  \label{TWEb}   \\
i\dot C &=& AB   \, ,  \label{TWEc}
\end{eqnarray}
These three equations
for the slowly-varying complex amplitudes $A$, $B$ and $C$ 
are the {\it three-wave equations}.
The relevance of these equations in various physical contexts is
discussed in \cite{H&L}.
They govern quadratic wave resonance in fluids and plasmas.
Their application to resonant Rossby wave triads
is considered in \cite{Lynch2003}.
In Appendix~A, we show that they are
a special case of the Nahm equations which are used to construct
soliton solutions in certain particle field theories.
For further references to the three-wave equations and a discussion
of their properties see \cite{Alber}.

The three-wave equations conserve the following three quantities:
\begin{eqnarray}
H &=& \half(ABC^*+A^*B^*C)  = \Re\{ABC^*\}
\,,\label{Hconstant} \\
N &=& |A|^2 + |B|^2 + 2|C|^2
\,,\label{Nconstant} \\
J &=& |A|^2 - |B|^2
\,.\label{Jconstant}
\end{eqnarray}
The equations are completely integrable. 
The following positive-definite combinations of $N$ and $J$ are
physically significant:
\[
N_{+} \equiv \half(N+J) = |A|^2 + |C|^2 \, , \qquad
N_{-} \equiv \half(N-J) = |B|^2 + |C|^2 \, .
\]
These combinations are known as the {\it Manley-Rowe relations}. 
Together with the Hamiltonian $H$, they provide three independent
constants of the motion.
We note that $H$ is invariant under the symmetry transformations
\begin{eqnarray}
(A,B,C) &\rightarrow& (A e^{i\chi},B e^{-i\chi},C) \,, \\
(A,B,C) &\rightarrow& (A e^{i\chi},B,C e^{i\chi}) \,, \\
(A,B,C) &\rightarrow& (A,B e^{i\chi},C e^{i\chi}) \,. 
\end{eqnarray}
These symmetries are associated, via Noether's theorem, with the 
three invariants $\big\{J, N_{+},N_{-}\big\}$.
Any two of the transformations generate the third.
This reflects the inter-dependence of $J$, $N_{+}$ and $N_{-}$.

\subsection{Reduction of the system}

To reduce the system for $H\ne0$, we express the amplitudes in polar form:
\begin{eqnarray}
A &=& |A|\exp(i\xi)       \label{TransA}
\,,\\
B &=& |B|\exp(i\eta )     \label{TransB}
\,,\\
C &=& |C|\exp(i\gamma)    \label{TransC}
\, .
\end{eqnarray}
In general, the phases of $A$, $B$ and $C$ are not periodic. However,
$\zeta = \gamma-(\xi+\eta)$ is periodic. 
The Hamiltonian may be written
\[
H = |A||B||C| \cos\zeta \,.
\]

The amplitude $|C|$ will be obtained in closed form in terms of
elliptic functions.  Once $|C|$ is known, $|A|$ and $|B|$ follow
immediately from the Manley-Rowe relations
\[
|A| = \sqrt{N_{+}-|C|^2} \,, \qquad |B| = \sqrt{N_{-}-|C|^2} \, .
\]
The phases $\xi$ and $\eta$ may now be determined. Using the three-wave
equations  (\ref{TWEa})--(\ref{TWEc}) together with equations
(\ref{TransA})--(\ref{TransC}), we find
\begin{equation}
\dot\xi = -\frac{ H}{ |A|^2} \, , \qquad \dot\eta = -\frac{ H }{ |B|^2} \, ,
\label{xieta}
\end{equation}
so that $\xi$ and $\eta$ can be obtained by quadratures.
Finally, $\zeta$ is determined unambiguously by
\begin{equation}
\frac{d|C|^2}{ dt}
= -2H\tan\zeta
\qquad {\rm and} \qquad H = |A||B||C|\cos\zeta \,.
\label{zeta}
\end{equation}
It also follows from (\ref{xieta}) and (\ref{zeta}) that
\begin{equation}
\dot\zeta = {H}\left(\frac{1}{|A|^2}+\frac{1}{|B|^2}-\frac{1}{|C|^2}\right) \,.
\label{zetadot}
\end{equation}
The phase of $C$ follows immediately, $\gamma=\xi+\eta+\zeta$, and
we can then reconstruct the complete solution using
(\ref{TransA})--(\ref{TransC}). 

\subsection{The equation for $|C|^2$}

%
\Figure{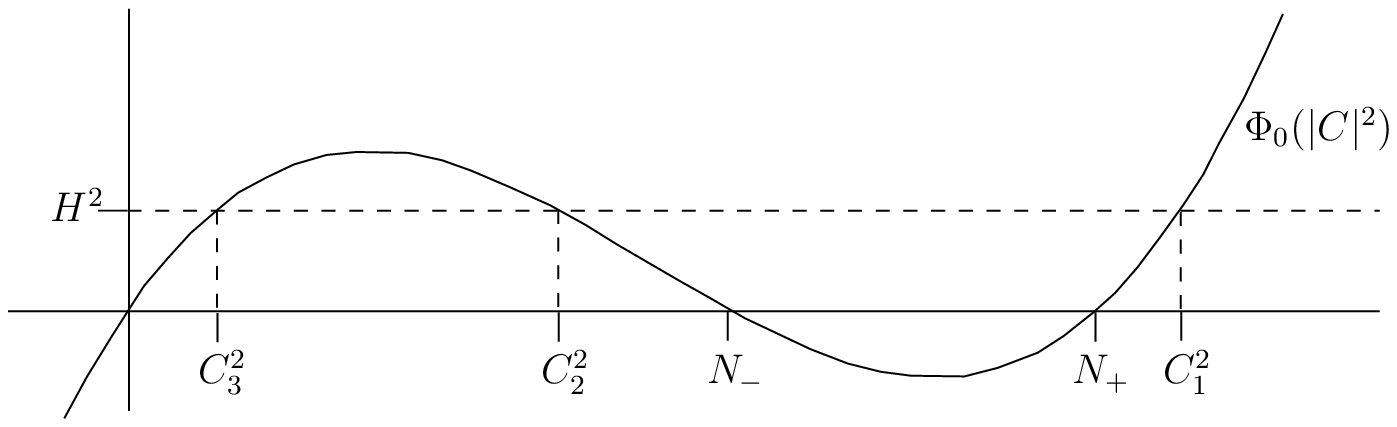}{0.80}{Fig:Phipoly}
{Polynomial $\Phi_0$ as a function of $|C|^2$.}
%

From the equation (\ref{TWEc}) for $\dot{C}$, and its complex conjugate
we get
\begin{equation}
\frac{d|C|^2}{dt} = 2\Im\{ABC^*\} \,.
\label{dccdt}
\end{equation}
Using the definition of the Hamiltonian, it follows that
\[
|A|^2 |B|^2 |C|^2 = H^2 + [\Im\{ABC^*\}]^2 \,.
\]
Applying this to the square of (\ref{dccdt}) and using the definitions of the
Manley-Rowe quantities immediately yields an equation for $|C|^2$ alone:
\begin{equation}
\left(\frac{d|C|^2}{dt}\right)^2 
= 4\Big[ (N_{+}-|C|^2)(N_{-}-|C|^2)|C|^2 - H^2 \Big] \,.
\label{dcsqdt}
\end{equation}
We define the cubic polynomial $\Phi_0(|C|^2)$
(plotted in Fig.~\ref{Fig:Phipoly}) by
\begin{equation}
\Phi_0(|C|^2) = (N_{+}-|C|^2)(N_{-}-|C|^2)|C|^2 \,.
\label{defPhi0}
\end{equation}
Then the right hand side of (\ref{dcsqdt}) may be written
$4[\Phi_0(|C|^2)-H^2]$.
For small $H^2$, this cubic has three positive
real roots.  If these roots, in descending order of magnitude, are
denoted $C_1^2$, $C_2^2$ and $C_3^2$, it follows that
\begin{equation}
0 \le C_3^2 \le C_2^2 \le N_{-} \le \half N 
\le N_{+} \le C_1^2 \le N \,. 
\label{Order1}
\end{equation}
(We have assumed without loss of generality that $J\ge0$).
In the case of equality of roots, the solution may be
obtained in terms of elementary functions. We assume in general that
this is not so and solve for $|C|^2$ in terms of
elliptic functions. However, before doing this, we investigate
perturbation motion about steady solutions.

\section{Small-Amplitude Modulation of Steady States}
\label{Sec:Perturbations}

We consider the case where the variations of the amplitudes
about their mean values are small. This enables us to make additional
approximations and derive simple estimates of the pulsation
period and rate of precession. From these two quantities, the
precession angle follows immediately.

\subsection{Steady State Motion}

We first consider solutions for which the amplitudes $|A|$, $|B|$ 
and $|C|$ are constant. The simplest cases are where the phases 
are also constant; then the three-wave equations become
\[
B^*C = CA^* = AB = 0 \,,
\]
which give three particular solutions
\begin{eqnarray*}
(i)   \qquad A = A_0 \,, \quad && B = C = 0 \,; \\
(ii)  \qquad B = B_0 \,, \quad && C = A = 0 \,; \\
(iii) \qquad C = C_0 \,, \quad && A = B = 0 \,.
\end{eqnarray*}
The first two solutions correspond to conical motions:
the bob moves in a circle, clockwise or anti-clockwise,
while the spring traces out a cone. These solutions are stable to small 
perturbations. The third particular case represents purely
vertical oscillations; this motion is unstable \cite{Lynch2002a}.

More generally, from (\ref{zeta}), constancy of the amplitudes implies 
$\zeta=\gamma-(\xi+\eta)=0$ so that $H=|A||B||C|$, and
the three-wave equations become
\begin{eqnarray}
-|A|\dot\xi      &=&   |B||C|  \nonumber    \\
-|B|\dot\eta     &=&   |C||A|  \label{twe0} \\ 
-|C|\dot\gamma   &=&   |A||B|  \nonumber    \,.   
\end{eqnarray}
Differentiating (\ref{dcsqdt}), a simple algebraic manipulation yields
\begin{equation}
|C|^2 = C_0^2 \equiv
\frac{1}{6} \left( 2N - \sqrt{N^2+3J^2} \right)
\label{C0sq}
\end{equation}
The other amplitudes are given by
\begin{eqnarray*}
|A|^2 &=& A_0^2 \equiv \frac{1}{6}
\left[ (N+3J) + \sqrt{N^2+3J^2} \right] \\
|B|^2 &=& B_0^2 \equiv  \frac{1}{6}
\left[ (N-3J) + \sqrt{N^2+3J^2} \right] \\
\end{eqnarray*}
These solutions are the elliptic-parabolic modes (EP-modes)
studied by Lynch \cite{Lynch2002a}. The precession rate is given by
$\Omega \equiv \dot\phi = \half(\dot\xi-\dot\eta)$ \cite{H&L}. 
From (\ref{twe0}) it follows that
\begin{equation}
\Omega = \frac{J C_0^2}{2H} \,.
\label{bigom}
\end{equation}
For $J=0$ we have planar motion with
\[
|C|^2 = \frac{N}{6} \,, \qquad |A|^2 = |B|^2 = \frac{N}{3} \,.
\]
These are the cup-like and cap-like solutions of Vitt and Gorelik
\cite{V&G}.

\subsection{Perturbation about Elliptic-Parabolic Motion}

We consider small deviations about the steady EP-mode solutions. We write
$|C|^2=C_0^2+\epsilon$ where $C_0^2$ is given by (\ref{C0sq}) and
 $|\epsilon| \ll C_0^2$. Then, if (\ref{dcsqdt}) is differentiated
and nonlinear terms in $\epsilon$ are omitted, we obtain
\begin{equation}
\frac{d^2 \epsilon}{dt^2}  
+ \left( 2\sqrt{N^2+3J^2} \right) \epsilon = 0 \,.
\end{equation}
The solution is $\epsilon(t) = \epsilon(0)\cos \omega_{\rm P}t$,
an oscillation about $C_0^2$ with the {\it pulsation frequency}
\begin{equation}
\omega_{\rm P} = \sqrt{2}{\root 4\of{N^2+3J^2}} \,.
\label{pulsefreq}
\end{equation}
For the EP-modes, the horizontal projection is an ellipse precessing 
at a constant rate $\Omega$. 
The perturbation is a pulsating motion, with sinusoidal time variation,
in which the major and minor axes of the ellipse
alternately expand and contract with period $T_{\rm P} =
2\pi/\omega_{\rm P}$. The area of the ellipse is proportional to $J$ and
remains constant \cite{H&L}. It is straightforward to derive expressions
in terms of elementary functions for the remaining amplitudes and the
phases, but they are not required to determine the precession angle.

We note from (\ref{pulsefreq}) that
$\sqrt{2N} \le \omega_{\rm P} \le 2\sqrt{2N}$. 
From the precession rate and the pulsation frequency, the precession angle
follows immediately:
\[
\Delta\phi = \Omega T_{\rm P} \,.
\]
Using (\ref{C0sq}), (\ref{bigom}) and (\ref{pulsefreq}), this gives us
\begin{equation}
\displaystyle{
\Delta\phi = 
\frac{J C_0^2}{2H} \frac{2\pi}{\omega_{\rm P}} =
\frac{\pi}{3} \left(\frac{J}{\sqrt{8}H}\right)
\left[\frac{2N-\sqrt{N^2+3J^2}}{\root 4\of{N^2+3J^2}}\right] \,. }
\label{delphiEP}
\end{equation}
For small angular momentum $J\ll N$, the term in square brackets is
close to $\sqrt{N}$ and
\begin{equation}
\Delta\phi \approx
\frac{\pi}{3} \left(\frac{J\sqrt{N}}{\sqrt{8}H}\right) \,.
\label{delphiEP1}
\end{equation}

\section{Analytical Solution of the Three-wave Equations}
\label{Sec:Analytical}

\subsection{Solution in Weierstrass Elliptic Functions}

We now derive an explicit analytical solution for $|C|^2$, valid for finite
amplitudes. The solutions for $|A|^2$ and $|B|^2$
follow immediately from the Manley-Rowe relations. Then
(\ref{xieta}) are integrated for the phases. The integrals
turn out to be similar to those occurring for the spherical pendulum,
so the approach of Whittaker \cite{Whittaker37} applies.
The required properties of the Weierstrass elliptic functions are
given in Whittaker and Watson \cite{WW}, Ch.~20 and in Lawden
\cite{Lawden}, Ch.~6 (see also
Abramowitz and Stegun \cite{A&S}, Gradshteyn and
Ryzhik \cite{Gradshteyn} and Byrd and Friedman \cite{Byrd}).

\subsubsection{Solution for the amplitudes}

The quadratic term on the right of (\ref{dcsqdt}) is removed by a simple
transformation $u=|C|^2/N-1/3$ and $\tau=\sqrt{N} t$.
Then we obtains
\begin{eqnarray}
\left(\frac{d\,u}{d\tau}\right)^2 
&=& 4u^3 - g_2 u - g_3  \nonumber \\
&=& 4 (u-e_1)(u-e_2)(u-e_3) \,.
\label{Weier}
\end{eqnarray}
This is the standard form of the equation for Weierstrass elliptic
functions. The constants $g_2$ and $g_3$,
called the {\it invariants}, are given by
\[
g_2 = \left( \frac{1}{3} + \frac{J^2}{N^2} \right)  \,, \qquad
g_3 = \left(
-\frac{1}{27}+\frac{J^2}{3N^2} + \frac{4H^2}{N^3} \right) \,.
\]
For small $H^2$, the discriminant $\Delta = g_2^3-27g_3^2$ is
positive and the three roots are real.
This is the case of physical interest, and
we assume the roots of the cubic are ordered so
that $e_1 > e_2 > e_3$.  Note that $e_1+e_2+e_3 = 0$. 
The general solution of (\ref{Weier}) is 
\[
u = \wp(\tau+\alpha)
\]
where $\alpha$ is an arbitrary (complex) constant.
The function $\wp(z)$ is defined by
\begin{equation}
\wp(z) = \frac{1}{z^2} + {\sum_{m,n}}^\prime \left\{
\frac{1}{(z-2m\omega_1-2n\omega_2)^2} -\frac{1}{(2m\omega_1+2n\omega_2)^2}
\right\}
\label{Pfcn}
\end{equation}
where the summation is over all integral $m$, $n$ except $m=n=0$. It has
poles on the real line and is doubly periodic: 
$\wp(z+2m\omega_1+2n\omega_2)=\wp(z)$ for all integers $m$ and $n$. 
The difficult problem of determining $\omega_1$ and $\omega_2$ from the invariants
is discussed in \S21.73 of \cite{WW}. The quantity $\omega_3$ is defined
by requiring $\omega_1+\omega_2+\omega_3=0$. It may be shown that
\begin{equation}
\wp(\omega_1) = e_1 \, \qquad 
\wp(\omega_2) = e_2 \, \qquad 
\wp(\omega_3) = e_3 \,.  
\label{pompom}
\end{equation}
In the present case,
$\omega_1$ is real and $\omega_3$ is pure imaginary
(explicit expressions are given below). On the real line,
$\wp(z)$ is real, with values in the range $[e_1,+\infty)$. On the line
$z=\omega_3+x$ it takes real values in the interval $[e_3,e_2]$.
Moreover, as $z$ varies along the edge of the rectangle
from $0$ to $\omega_1$ to $\omega_1+\omega_3(=-\omega_2)$
to $\omega_3$ to $0$,
$\wp(z)$ is real and decreases monotonically from $+\infty$ to
$e_1$ to $e_2$ to $e_3$ to $-\infty$.
To satisfy the initial conditions, we choose
$\alpha=\omega_3-\tau_0$, where $\tau_0$ is real and may be taken
as zero by a suitable choice of time origin.
Then $\wp(\tau+\omega_3)$ is real and oscillates
between $e_3$ and $e_2$.
The solution for the amplitude is
\begin{equation}
|C|^2 = N\left[ \frak{1}{3} + \wp(\tau+\omega_3) \right] \,.
\label{Wsoln}
\end{equation}
The behaviour of the Weierstrass $\wp$-function is shown in
Figure~\ref{weiersurf}.  For general $z$ it takes complex values.
On the line $\Im\{z\}=\omega_3$ the function
is real with periodic oscillations, as indicated by the heavy lines
at the front of the figure. 


\Figure{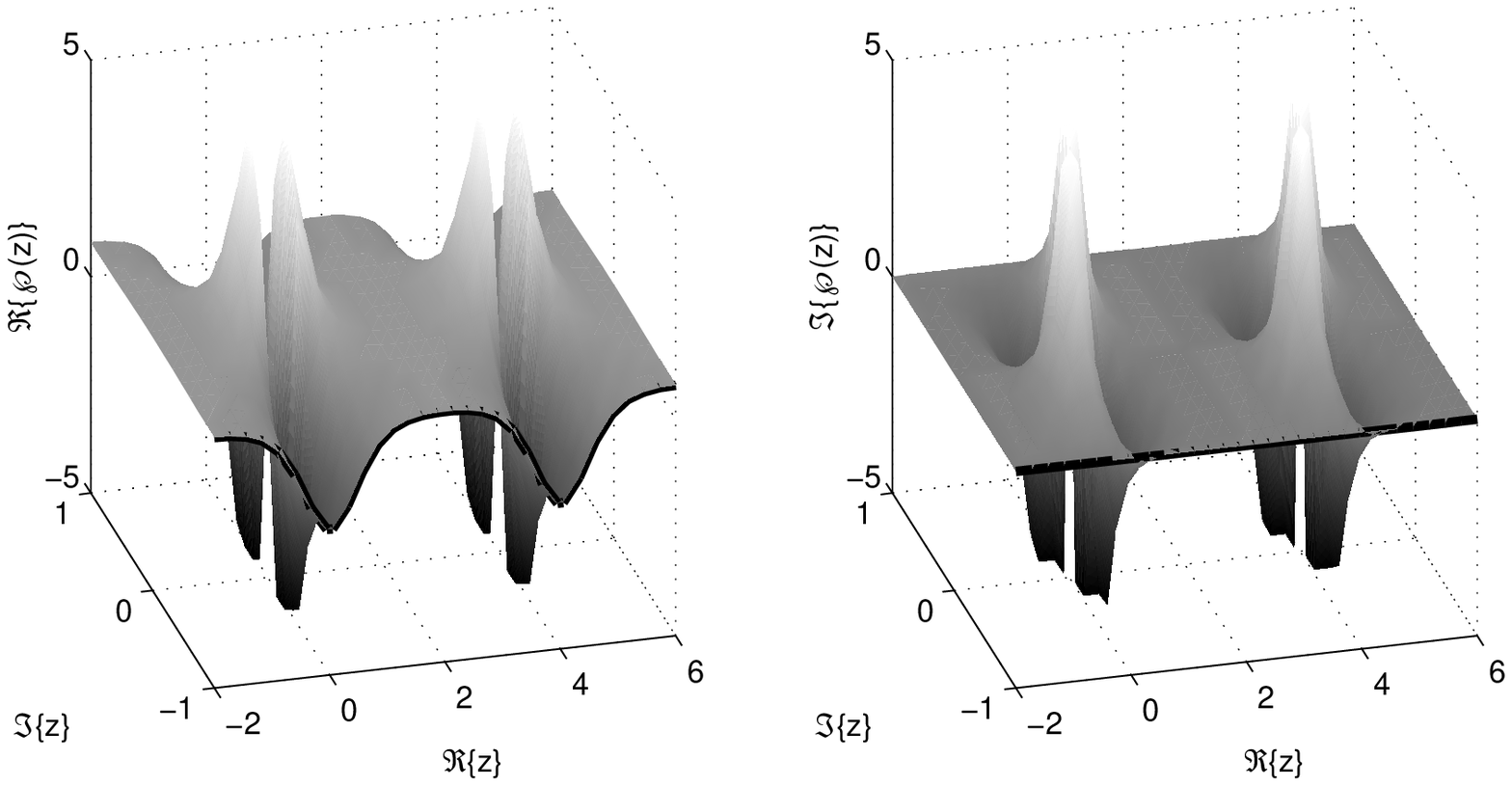}{0.80}{weiersurf}{Weierstrass's $\wp$-function 
with half-periods $\omega_1=2$, $\omega_3=i$ on the domain
$\{z=x+iy: x\in[-2,+6], y\in[-1,+1]$. Left panel: real part; 
right panel: imaginary part. Values for $z=x-\omega_3$ are plotted as
heavy lines. Calculations are based on (\protect\ref{Pfcn}). The function has
double poles at $z=2m\omega_1+2n\omega_2$.}


\subsubsection{Solution for the phase angles}

Weierstrass's zeta function is defined by
\begin{equation}
\frac{d\mathbf{\zeta}}{dz} = - \wp(z) \,, \qquad
\lim_{z\rightarrow0}\, [\mathbf{\zeta}(z)-z^{-1}] = 0 \,.
\label{zetadef}
\end{equation}
It is quasi-periodic in the sense that
\begin{equation}
\mathbf{\zeta}(z+2\omega_1) = 
\mathbf{\zeta}(z) + 2\mathbf{\zeta}(\omega_1) \,.
\label{zetaper}
\end{equation}
We note that $\mathbf{\zeta}(z)$ is an odd function of $z$ and will
use the relation
\begin{equation}
\omega_1 \mathbf{\zeta}(\omega_2) -\omega_2 \mathbf{\zeta}(\omega_1)
= \half\pi i \,.
\label{ozoz}
\end{equation}
The sigma function is defined by
\begin{equation}
\frac{d}{dz}\log\sigma(z) =  \mathbf{\zeta}(z) \,, \qquad
\lim_{z\rightarrow0}\, \frac{\sigma(z)}{z} = 1 \,.
\label{sigmadef}
\end{equation}
It is also quasi-periodic, such that
\begin{equation}
\sigma(z+2\omega_1) =
-\exp[2\mathbf{\zeta}(\omega_1)(z+\omega_1)]\,\sigma(z) \,.
\label{sigmaper}
\end{equation}
Three other sigma functions may be defined. The relationship between
the sigma functions and the Weierstrass $\wp$-function is similar to
that between the theta functions and the Jacobi elliptic
functions. We will not require these.  We will require the identity
\begin{equation}
\frac{\wp^\prime(\alpha)}{\wp(z)-\wp(\alpha)} = 
\mathbf{\zeta}(z-\alpha)-\mathbf{\zeta}(z+\alpha)+
                        2\mathbf{\zeta}(\alpha) 
\label{ppzz}
\end{equation}
(this follows from a consideration of the poles and zeros of the
functions on each side).

The solution (\ref{Wsoln}) leads to a solution for $|A|^2$:
\begin{equation}
|A|^2 =  \frac{N+3J}{6}-N\wp(\tau+\omega_3) \,.
\label{Asoln}
\end{equation}
Substituting in the first of (\ref{xieta}) we have
\[
\sqrt{N}\,\frac{d\xi}{d\tau} = 
\frac{6H}{6N\wp(\tau+\omega_3)- (N+3J)}
\]
Now we introduce auxiliary constants $\kappa_{\pm}$ defined by
\[
\wp(\kappa_{+}) = \frac{N+3J}{6N} \equiv e_{+} \,, \qquad
\wp(\kappa_{-}) = \frac{N-3J}{6N} \equiv e_{-} \,.
\]
Using (\ref{Weier}), it follows that
\[
[\wp^\prime(\kappa_{+})]^2 = [\wp^\prime(\kappa_{-})]^2 = 
-\left(\frac{4H^2}{N^3} \right) \,.
\]
We must determine which sign for the derivatives should be chosen.
From (\ref{Order1}) the following sequence of inequalities holds:
\begin{equation}
-\frak{1}{3} \le e_3 \le e_2 \le e_{-} 
\le \frak{1}{6} \le e_{+} \le e_1 \le \frak{2}{3} \,. 
\label{Order2}
\end{equation}
Since $e_2 < e_{-} < e_{+} < e_1$, it follows that
$\kappa_{\pm}$ lie on the line between $\omega_1$ and
$\omega_1+\omega_3$, which determines the sign of the derivatives
to be $\wp^\prime(\kappa_{\pm}) = 2iH/N^{3/2}$, a positive imaginary number.
The equation for $\xi$ thus becomes
\[
\frac{d\xi}{d\tau} = \left(\frac{1}{2i}\right)
\frac{\wp^\prime(\kappa_{+})}{\wp(\tau+\omega_3)-\wp(\kappa_{+})}
\]
Using (\ref{ppzz}) this may be expressed in terms of zeta functions
and using (\ref{sigmadef}) it may be integrated immediately to yield
\begin{equation}
\xi - \xi_0 =   \left(\frac{1}{2i}\right)
\left\{
\log\left[ \frac{\sigma(\tau+\omega_3-\kappa_{+})}
               {\sigma(\tau+\omega_3+\kappa_{+})} \right]
+ 2 \mathbf{\zeta}(\kappa_{+}) \tau
\right\} \,.
\end{equation}
A similar expression holds for $\eta-\eta_0$ with $\kappa_{-}$
replacing $\kappa_{+}$. Thus we obtain the expression for the
azimuthal angle $\phi$:
\[
{\phi - \phi_0}  = 
\left( \frac{1}{2i} \right)
\left\{
[\mathbf{\zeta}(\kappa_{+})-\mathbf{\zeta}(\kappa_{-})] \tau
+\frac{1}{2}\log\left[ \frac{\sigma(\tau+\omega_3-\kappa_{+})}
               {\sigma(\tau+\omega_3+\kappa_{+})}
       \frac{\sigma(\tau+\omega_3+\kappa_{-})}
               {\sigma(\tau+\omega_3-\kappa_{-})} \right] 
\right\} \,.
\]
This is the solution for the azimuth as a function of time.
Using the quasi-periodic properties (\ref{zetaper}) and
(\ref{sigmaper}), the change in $\phi$ when $\tau$ varies
by $2\omega_1$ may be computed:
\begin{equation}
\fbox{$
\Delta\phi =
-i\omega_1\big(
\mathbf{\zeta}(\kappa_{+})-\mathbf{\zeta}(\kappa_{-})\big)
+ i\mathbf{\zeta}(\omega_1)(\kappa_{+}-\kappa_{-})
$} 
\label{delphi1}
\end{equation}
This is the desired analytical expression for the pulsation angle.%
\footnote{%
The apparent discrepancy with the result of Whittaker for the spherical
pendulum (p.~106 in \cite{Whittaker37})
arises from our choice of convention that
$\Im\{\omega_3/\omega_1\}>0$. Our result is consistent with the
rigid body formula (7.3.24) in Lawden \cite{Lawden}, who adopts the same
convention as we do.}

We note two obvious special cases of (\ref{delphi1}).
When $J = 0$ we have $\kappa_{+}=\kappa_{-}$,
yielding a zero result for $\Delta\phi$.
When $H=0$, we have $\kappa_{+}=\omega_1$ and $\kappa_{-}=-\omega_2$,
so
\begin{equation}
\Delta\phi =
-i [ \omega_1 \mathbf{\zeta}(\omega_2)
    -\omega_2 \mathbf{\zeta}(\omega_1) ] = \frac{\pi}{2} \,,
\label{dphihalfpi}
\end{equation}
where we have used (\ref{ozoz}). These two special cases
intersect in the homoclinic orbit (with $J=H=0$) which has an
infinite transition time.


\subsection{Solution in Jacobi Elliptic Functions}

While (\ref{delphi1}) is the analytical solution,
it is not immediately
obvious how numerical information may be extracted from it.
The quantities on the right side are all computable in principle,
but at the expense of considerable effort. It is therefore useful 
to seek an alternative expression, in terms of Jacobi elliptic
functions.

\subsubsection{Solution for the amplitudes}
 
Recall that with the transformation $u=|C|^2/N-1/3$ and
$\tau=\sqrt{N}t$, (\ref{dcsqdt}) was transformed to (\ref{Weier}),
which we write again for convenience:
\begin{equation}
\left(\frac{d\,u}{d\tau}\right)^2 
= 4 (u-e_1)(u-e_2)(u-e_3) \,.
\label{Weier2}
\end{equation}
For solutions of physical interest, $H^2$ is sufficiently small that
the three roots of the cubic are real.
Defining the quantities
\[
k^2 = \left( \frac{ e_2-e_3 } { e_1-e_3 } \right) 
\qquad \mbox{and} \qquad 
\nu^2 = ( e_1-e_3 )  \,,
\]
a further transformation,
\[
w = \sqrt{\frac{ u-e_3 }{ e_2-e_3 } } \,, \qquad
s = \nu\tau \,,
\]
brings equation (\ref{Weier2}) to the standard form
\begin{equation}
\left( \frac{dw}{ds} \right)^2 = (1-w^2)(1-k^2w^2) \,.
\label{dwds}
\end{equation}
The solution is $w=\sn(s-s_0)$, or 
\[
u = e_3 + (e_2-e_3)\,\sn^2(s-s_0)
\]
where $s_0$ is arbitrary.
The Jacobi elliptic function $\sn\,s$ has period $4K$, where 
\begin{equation}
K = K(k) = \int_0^1 \frac{dw}{\sqrt{(1-w^2)(1-k^2w^2)} }  \,,
\label{ComEllInt}
\end{equation}
so $\sn^2(s-s_0)$ has period $2K$.
For definiteness, we set $s_0=0$, which means choosing the origin of time
where the solution has a minimum:
\begin{equation}
|C|^2 =   C_3^2 + (C_2^2-C_3^2)\,\sn^2(\nu\sqrt{N} t)  \,.
\label{Jsoln}
\end{equation}
Clearly, $|C|$ oscillates between $C_3$ and $C_2$ with physical
period
\begin{equation}
T = 2K/\nu\sqrt{N} \,.
\label{Period-T}
\end{equation}
The remaining amplitudes, $|A|$ and $|B|$, follow from
the Manley-Rowe relations:
\[
|A|^2 = {N_{+}-|C|^2} \,, \qquad |B|^2 = {N_{-}-|C|^2} \, .
\]
They have the same period as $|C|$ but vary in anti-phase with it and in phase 
with each other. We denote the minimum and maximum values of $|A|$ by
$A_3$ and $A_2$, and similarly for $|B|$. Thus
\[
N_{+} = A_3^2 + C_2^2 = A_2^2 + C_3^2 \,, \qquad
N_{-} = B_3^2 + C_2^2 = B_2^2 + C_3^2 \,.
\]
The initial values of the amplitudes (for $s_0=0$) are
\[
|A(0)| = A_2 \,, \qquad |B(0)| = B_2 \,, \qquad |C(0)| = C_3 \,.
\]

We note here an important scaling invariance of the three-wave equations.
If the amplitudes are magnified by a constant factor and the time is
contracted by the same factor, the form of the equations
(\ref{TWEa})--(\ref{TWEc}) is unchanged.
Thus, the period of the modulation envelope motion varies inversely with
its amplitude. The overall scale may be measured by $\sqrt{N}$ and
the inverse dependence of $T$ on this is seen in (\ref{Period-T}).

The solutions (\ref{Wsoln}) and (\ref{Jsoln}) must be equivalent.
This follows from identities relating Weierstrass and Jacobi
elliptic functions.  The complimentary modulus is defined as
$k^\prime = \sqrt{1-k^2}$, and we write $K^\prime=K(k^\prime)$.
The parameters are related by
\[
k=\sqrt{\frac{e_2-e_3}{e_1-e_3}} \,, \quad
k^\prime=\sqrt{\frac{e_1-e_2}{e_1-e_3}} \,, \quad
\omega_1=\frac{K}{\sqrt{e_1-e_3}} \,, \quad
\omega_3=\frac{iK^\prime}{\sqrt{e_1-e_3}}  
\]
(\cite{Gradshteyn}, p.~919). Then we have
\[
\wp(z) = e_3+\frac{e_1-e_3}{\sn^2(\sqrt{e_1-e_3}\,z)}
\]
But the Jacobi function $\sn(s+iK^\prime)$ is given in terms of its
value on the real line by
\[
\sn(s+iK^\prime) = \frac{1}{k\,\sn\, s}
\]
and the equivalence between the two forms of solution follows
immediately.

\subsubsection{Solution for the phase angles}

It remains to determine the phases.
Integration of (\ref{xieta}) furnishes
the angles $\xi$ and $\eta$. We define
\begin{eqnarray*}
\gamma_{+}^2 =\frac{C_2^2-C_3^2}{N_{+}-C_3^2}
             = \frac{e_2-e_3}{e_{+}-e_3} \,,
&\mbox{\quad}&
\lambda_{+}=\frac{H}{\nu\sqrt{N}{A_2^2}} = 
\frac{ H/N^{3/2} }{\sqrt{e_1-e_3}(e_{+}-e_3)}  \,,
\\
\gamma_{-}^2 =\frac{C_2^2-C_3^2}{N_{-}-C_3^2}
             = \frac{e_2-e_3}{e_{-}-e_3} \,,
&\mbox{\quad}&
\lambda_{-}=\frac{H}{\nu\sqrt{N}{B_2^2}} =
\frac{ H/N^{3/2} }{\sqrt{e_1-e_3}(e_{-}-e_3)}  \,.
\end{eqnarray*}
It follows from (\ref{Order2}) that $k^2<\gamma_{+}^2<\gamma_{-}^2<1$.
We may now write (\ref{xieta}) in the form
\begin{equation}
\frac{d\xi}{ds} = -\frac{\lambda_{+}}{1-\gamma_{+}^2 \,\sn^2 s}
\,,  \qquad
\frac{d\eta}{ds} = -\frac{\lambda_{-}}{1-\gamma_{-}^2 \,\sn^2 s} \,.  
\label{xieta2}
\end{equation}
The right sides are the integrands occuring in Legendre's
elliptic integral of the third kind (\cite{A&S}, p.~590).
They may be put in standard algebraic form by
defining $x=\sn\,s$. Writing
\[
\Pi(s,a,k) \equiv  \int_0^s \frac{ds}{1-a\, \sn^2 s } =
\int_0^x \frac{dx}{(1-a x^2)\sqrt{(1-x^2)(1-k^2x^2)}} \,,
\] 
the solution for $\xi$ becomes
\begin{equation}
\xi - \xi_0 = - \lambda_{+} \Pi\big(s, \gamma_{+}^2,k \big) \,.
\label{xixi01}
\end{equation}
There is an analogous solution for $\eta$. 
The changes in $\xi$ and $\eta$ over a half period $s\in[0,K]$ are
\[
\half\Delta\xi =  -\lambda_{+} \Pi(\gamma_{+}^2,k) \,, \qquad
\half\Delta\eta = -\lambda_{-} \Pi(\gamma_{-}^2,k) \,,
\] 
where the {\it complete}\/ elliptic integral is defined as
$\Pi(a,k)=\Pi(K,a,k)$.
The azimuthal angle of the pendulum is $\phi = \half(\xi-\eta)$.
Thus, the change in the azimuth over a full pulsation period is
\begin{equation}
\fbox{$
\Delta\phi =
 - \Big( \lambda_{+}\Pi(\gamma_{+}^2,k) - \lambda_{-}\Pi(\gamma_{-}^2,k) 
\Big) \,.
$} 
\label{dphi}
\end{equation}

In Appendix~B, an alternative formula (\ref{delphi3}) is derived from
the expression (\ref{dphi}), which is structurally
similar to (\ref{delphi1}) obtained above. Using this formula,
the limiting cases $\Delta\phi=0$ for $J=0$ and
$\Delta\phi=\pi/2$ for $H=0$ are again derived, in agreement 
with (\ref{dphihalfpi}).


\section{Approximate Formulas for the Precession Angle}
\label{Sec:Approximations}

We have derived an exact analytical expression 
for the precession angle, involving elliptic integrals.
It is of interest to obtain more convenient approximate formulas,
involving only elementary functions.
It might be expected that the easiest
way to do this would be to approximate (\ref{dphi}) directly.
However, it turns out that it is easier, and more transparent,
to return to the differential equations governing the the system,
use them to write down an integral for the precession angle and
approximate this integral.


The precession angle is $\phi=\half(\xi-\eta)$. Combining the two
components of (\ref{xieta}), we obtain
\begin{equation}
\frac{d\phi}{dt} = \frac{JH}{2|A|^2|B|^2} \, .
\label{phidot}
\end{equation}
Eq.~(\ref{dccdt}) may be written
\begin{equation}
\frac{d|C|^2}{dt} = \pm 2 \sqrt{ |A|^2 |B|^2 |C|^2 - H^2 } \,.
\label{dCCdt}
\end{equation}
Taking the quotient of these two equations, we get
\begin{equation}
\frac{d\phi}{d |C|^2} = 
\pm  \frac{JH}{4|A|^2|B|^2\sqrt{ |A|^2 |B|^2 |C|^2 - H^2 }} \,.
\label{dphidCC}
\end{equation}



The pulsation of the amplitude $|C|$ occurs 
between $C_2$ and $C_3$, where $C_3^2$ and $C_2^2$ are
the two smallest zeros of the polynomial
\begin{equation}
\Phi=|A|^2|B|^2|C|^2-H^2 =
|C|^2(|C|^2-N_{+})(|C|^2-N_{-}) - H^2 \,.
\end{equation}
It is also useful to write $\Phi=\Phi_0-H^2$ where
\[
\Phi_0=|A|^2|B|^2|C|^2 = |C|^2(|C|^2-N_{+})(|C|^2-N_{-}) 
\]
is as defined by (\ref{defPhi0}) and illustrated in Fig.~\ref{Fig:Phipoly}.
The two signs in the differential equation (\ref{dphidCC}) correspond 
to phase changes during alternate half-cycles of the pulsation.
The integral of (\ref{dphidCC}) over a full cycle may be written
formally: 
\begin{equation}
\Delta\phi=\frac{JH}{2}\int_{C_3^2}^{C_2^2}\frac{d|C|^2}{|A|^2|B|^2\sqrt{\Phi}}.
\label{Intc1c2}
\end{equation}
It is convenient to change the integration limits;
to do this, we consider (\ref{Intc1c2}) as an integral over the
complex $Z$-plane, where $Z=|C|^2$ on the positive real axis.
This gives 
\begin{equation}
\Delta\phi =
\frac{JH}{4}\int_{{\mathscr C}_1}\frac{dZ}
{(Z-N_{+})(Z-N_{-})\sqrt{\Phi(Z)}}.
\end{equation}
The contour ${\mathscr C}_1$ encircles $C_3^2$ and
$C_2^2$ and the square root in the integrand has two branch cuts, one
from $C_3^2$ to $C_2^2$ and the other from $C_1^2$ to $+\infty$.
This is illustrated in Fig.~\ref{Fig:Branches}. In addition to
the three branch points, the integrand has two simple poles
at $Z=N_{+}$ and $Z=N_{-}$. In fact, the residues at these two
poles sum up to zero:
\begin{equation}
\mbox{Res}(N_{+}) = -\mbox{Res}(N_{-}) =
\frac{-iJ}{4}\frac{1}{N_+-N_-} = -\frac{i}{4} \,.
\end{equation}
Furthermore, the integrand goes to zero sufficiently fast 
as $|Z|\rightarrow\infty$ that we can replace the contour
${\mathscr C}_1$ by ${\mathscr C}_2$ (Fig.~\ref{Fig:Branches}).
Returning to the original integral (\ref{Intc1c2}),
this corresponds to a change of integration range to 
\begin{equation}
\Delta\phi=\frac{JH}{2}\int^{\infty}_{C_1^2}
\frac{d|C|^2}{|A|^2|B|^2\sqrt{\Phi}} \,.
\label{AxIntCplus}
\end{equation}
This interval is more convenient than the previous one because the
integrand is small everywhere except near the lower limit of
integration and because the point of inflection in $\Phi$ can
cause difficulties when approximating $\Phi$ near $C_2^2$.
%
%
\Figure{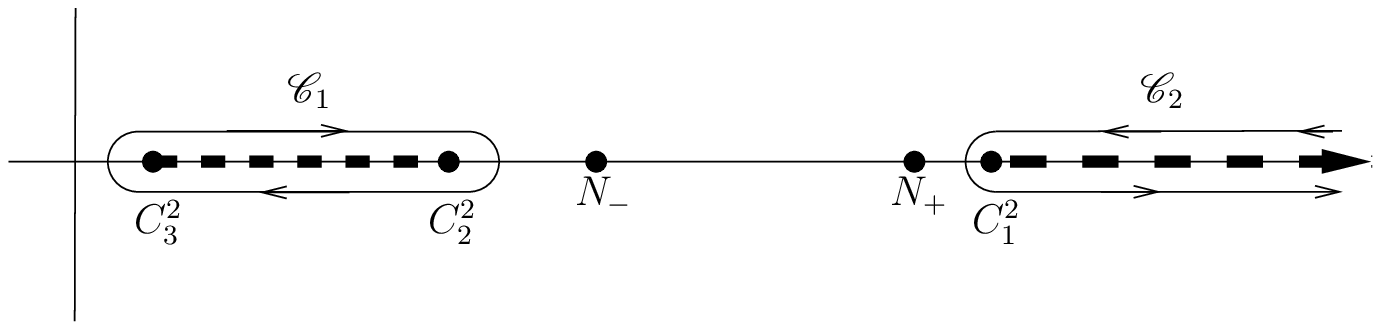}{0.80}{Fig:Branches}
{Contours $\mathscr{C}_1$ and $\mathscr{C}_2$ in the $Z$-plane.}
%
%
Since the integrand is dominated by its behaviour near $C_1^2$ the
obvious approach would be to find a quadratic which approximates $\Phi$
near this point. As $C_1^2$ is the root of a cubic, it
can be written in terms of $H$, $J$ and $N$, but this expression
is cumbersome and does not yield a convenient approximation.
It is simpler to consider the behaviour
of $\Phi$ at $N_{+} =(N+J)/2$. This point is
close to $C_1^2$ because $H^2$ must be small compared to $N$ for the
periodic motion to exist.%
\footnote{It can be shown easily that the maximum allowed value
of $H^2/N^3$ is $H_{00}^2=1/54\approx 0.0185$ and occurs for $J=0$.}

Having decided to approximate at $N_{+}$ rather than $C_1^2$,
the next step is to approximate $\Phi_0=Z(Z-N_{+})(Z-N_{-})$
by a quadratic with a root at $N_{+}$:
\begin{equation}
\Psi_0= Z_0 (Z-N_{+})(Z - Z_1) \,.
\label{GenPsi}
\end{equation}
It is possible to perform the resulting approximate integral.
However, the solution is complicated unless $Z_1=N_{-}$
(see Appendix~C).  Thus, we consider
\begin{equation}
\Psi_0=Z_0(Z-N_{+})(Z-N_{-}) \,.
\label{Quadpsi}
\end{equation}
The quadratic $\Psi_0$ and cubic $\Phi_0$ both vanish at $Z=N_{+}$ and
$Z=N_{-}$. They are also equal when $Z=Z_0$.
We consider two choices of $Z_0$.

First, we choose $Z_0$ to be the mean of $Z=N_{+}$ and $Z=N_{-}$,
that is $Z_0=N/2$.  The integral (\ref{AxIntCplus}) becomes
\begin{equation}
\Delta\phi=\frac{JH}{2}\int^{\infty}_{Z_{+}}
\frac{d Z}{(Z-N_{+})(Z-N_{-})\sqrt{\Psi_0(Z)-H^2}} \,.
\label{AxIntB}
\end{equation}
where $Z_{+}$ is the larger root of $\Psi_0-H^2=0$.
Defining $\sigma=2Z-N$, we get
\begin{equation}
\Delta\phi = \int_{\sigma_{+}}^{+\infty} 
\frac{2\sqrt{2} JH d\sigma }{(\sigma^2-J^2)\sqrt{\sigma^2-(J^2+8H^2/N)}} \,,
\label{dphidCCapproxx}
\end{equation} 
where $\sigma_{+}=\sqrt{J^2+8H^2/N}$.
This may be integrated analytically (\cite{Dwight}, p.~72) to give
\begin{equation} 
\Delta\phi = - \tan^{-1}\left\{
\textstyle{\left( \frac{\sqrt{8}H}{\sqrt{N}J}\right)
\frac{\sigma}{\sqrt{\sigma^2-(J^2+8H^2/N)}} }
\right\} \bigg|_{\sigma_{+}}^{+\infty}
=
\left[
 \frac{\pi}{2} - \tan^{-1} \left( \frac{\sqrt{8}H}{\sqrt{N}J} \right)\right]
\end{equation} 
Noting that $\tan^{-1} x = \frac{\pi}{2}-\tan^{-1}\frac{1}{x}$,
the phase change over a full cycle is
\begin{equation}
\Delta\phi \approx 
\tan^{-1}\displaystyle{\left(\frac{\sqrt{N}J}{\sqrt{8}H}\right)} \,.
\label{DulCush}
\end{equation}
This elegant approximate formula for the pulsation angle was reported
by Dullin, Giacobbe and Cushman \cite{Dullin}, although they did not
give the factor $\sqrt{8}$ explicitly, and we refer to it as the DGC
formula.  Numerical experiments indicate that it is of high accuracy
throughout the accessible domain.

An alternative choice of quadratic approximation requires
$\Psi_0$ and $\Phi_0$ to have equal derivatives
at $Z=N_+$. In this case $Z_0=N_{+}$. We integrate, again taking the
lower limit to be the larger root of $\Psi_0-H^2=0$, to get
\begin{equation}
\Delta\phi \approx \tan^{-1}\left(\frac{J\sqrt{N+J}}{\sqrt{8}H}\right)
\label{Altform}
\end{equation}
It will be shown below that this formula is also in resonable agreement
with the analytical solution.

The above approximations are subtle: we replace a cubic by a quadratic,
changing the integrand, but we also change the lower limit. These
effects tend to compensate, resulting in surprisingly accurate
approximations.  Moreover, it is found
that the two approximations (\ref{DulCush}) and (\ref{Altform})
have errors which are of opposite sign and approximately equal.
Choosing $Z_0=(N+\alpha J)/2$ in (\ref{Quadpsi}), we get the 
approximation 
\begin{equation}
\fbox{$
\Delta\phi \approx \tan^{-1}\left( \displaystyle{
\frac{J\sqrt{N+\alpha J}}{\sqrt{8}H}  } \right)
$} 
\label{Optform}
\end{equation}
By numerical experiment, we deduce the optimal value $\alpha=0.458$.
Numerical results using the various approximations will
be presented in the following section and (\ref{Optform})
will be found to yield remarkably accurate results.


\section{Numerical Experiments}
\label{Sec:Numerics}

We first compare the precession angle calculated using
the exact analytical expression (\ref{dphi}) with values extracted from a
numerical integration of the three-wave equations 
(\ref{TWEa})--(\ref{TWEc}). 
For given $N$ and $J$, the maximum value of the cubic $\Phi_0(Z)$
is at $Z_{\rm max} =\frac{1}{6}[2N-\sqrt{N^2+3J^2}]$. Thus, the
maximum value of $H$ is
\[
H_{\rm 0} = H_{\rm 0}(N,J) = \sqrt{\Phi_0(Z_{\rm max})} \,.
\]
The three-wave equations were solved for a range of values
$0 \le J \le 1$ and $H$ covering the accessible parameter domain
$0 \le H \le H_{\rm 0}$.
We take $N=1$ in all cases; this is no loss of generality, as it is
equivalent to a rescaling of the amplitudes by $N^{-1/2}$ and of the
time by $N^{1/2}$.  From the numerical solution, the major and minor axes
\[
A_{\rm maj} = |A| + |B| \qquad \hbox{and} \qquad
A_{\rm min} = |A| - |B|
\]
of the osculating or instantaneous ellipse (see \cite{H&L})
were calculated as functions of time, and the precession angle
was evaluated as the change in $\phi$ between
successive maxima of $A_{\rm maj}$.
The precession angle was computed as a function of $J$ and $H$.
The results are presented in Fig.~\ref{Fig:NumAna}.
The heavy line is $H_{\rm 0}(J)$.
The left-hand panel shows $\Delta\phi$
calculated using the analytical formula (\ref{dphi}).
The precession angle vanishes for $J=0$ and is equal to $90^\circ$
for $H=0$.
The center panel shows the angle calculated from the numerical
solution of the three-wave equations. It is very similar to
the analytical result.
The right-hand panel shows the
difference between the precession angle
calculated from the numerical solution and the analytical
formula. The values are generally very small (the contour interval
in Fig.~\ref{Fig:NumAna}(C) is $0.1^\circ$).
The maximum difference is $0.6^\circ$ and the
discrepancy may be ascribed to numerical noise. 


\Figure{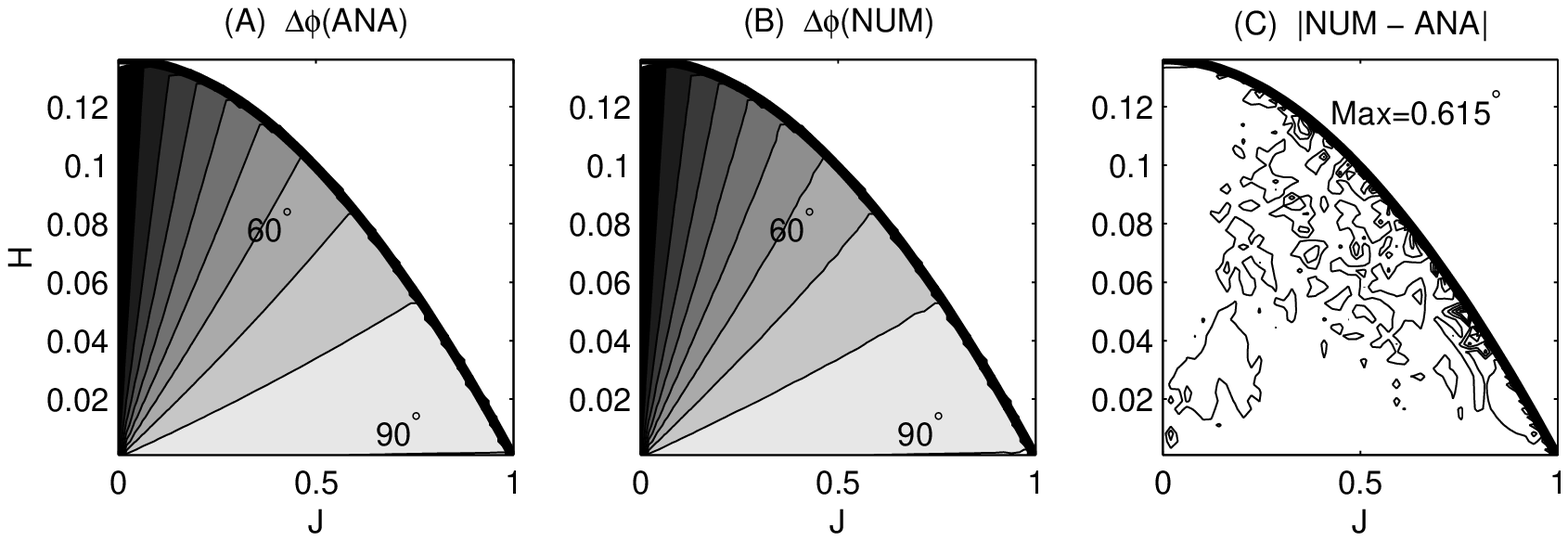}{0.80}{Fig:NumAna}
{Left-hand panel: precession angle $\Delta\phi$ calculated using
the exact analytical formula (\protect\ref{dphi}).
Center panel: $\Delta\phi$ calculated from numerical integration
of the three-wave equations. 
Note that $\Delta\phi=0$ for $J=0$ and $\Delta\phi=90^\circ$
for $H=0$. 
Right-hand panel: difference in precession angle
between the numerical and the analytical solution.}


\subsection{Determination of the Precession Angle}

We now show that the envelope of the motion may be determined
to high accuracy by using approximate formulas involving
only elementary functions.
We use the analytical values as a reference to evaluate
the accuracy of the approximate formulas.
In Fig.~\ref{Fig:DgcAna} the differences between the exact and
approximate expressions for $\Delta\phi$ are shown. 
The absolute values of these errors are plotted.
The maximum error in the DGC formula (Fig.~\ref{Fig:DgcAna}(A))
is about $2.2^\circ$, and occurs for $J\approx\half$ 
and $H$ at its maximum permissible value. 
The error in the alternative formula (\ref{Altform}) is of comparable 
magnitude, with a maximum of about $2.5^\circ$ (Fig.~\ref{Fig:DgcAna}(B)),
but is of opposite sign. 
The optimal value $\alpha=0.458$ of the parameter in the formula
(\ref{Optform}) was found by experiment.
Fig.~\ref{Fig:DgcAna}(C) shows that this formula is
significantly more accurate, with a maximum error less than $0.4^\circ$.
This is a remarkable level of precision, considering the simplicity of the
formula.  The compensation of errors leads to
what might be described as the {\it unreasonable effectiveness}\/
of the approximation. 


\Figure{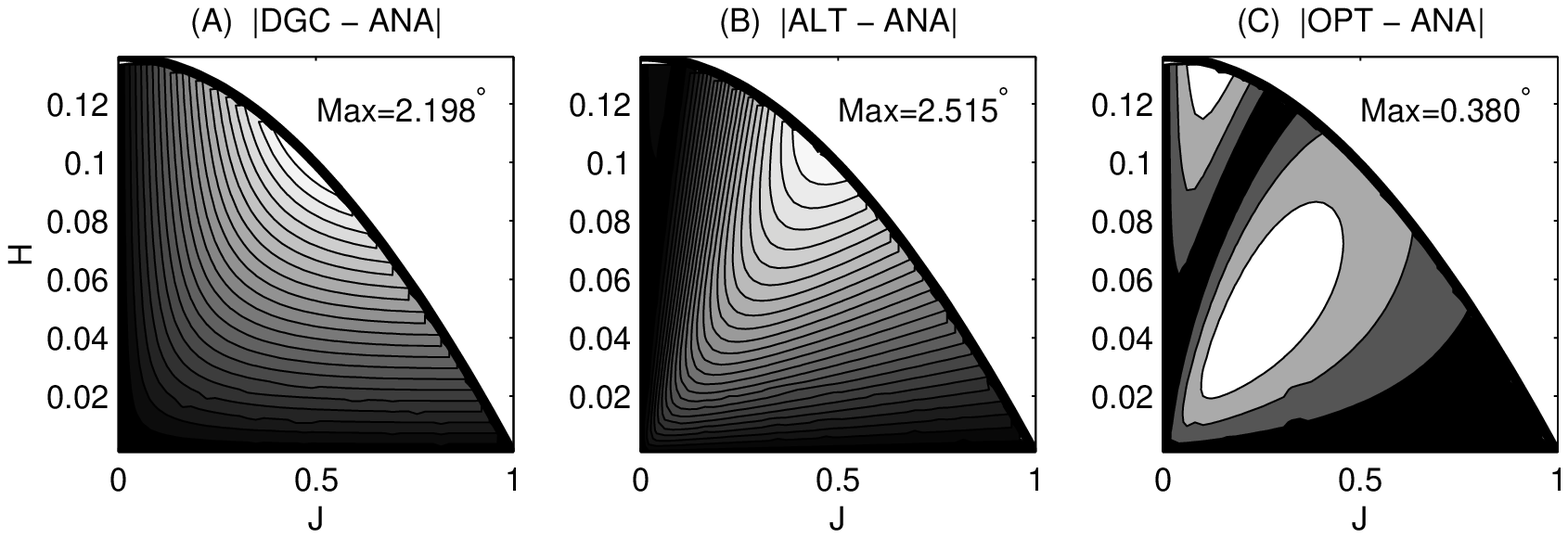}{0.80}{Fig:DgcAna}
{Differences in precession angle $\Delta\phi$ between three
approximate formulas and the analytical solution (\ref{dphi}).
(A) The DGC formula (\protect\ref{DulCush});
(B) The alternative formula (\protect\ref{Altform});
(C) The optimum formula (\protect\ref{Optform}).
Absolute values are shown. The signs of the errors of
(\protect\ref{DulCush}) and (\protect\ref{Altform}) are opposite.
The contour interval is $ 0.1^\circ$ in all panels.}


\subsection{Determination of the Pulsation Amplitude}

The extent to which energy is exchanged between
the elastic and pendular modes of oscillation may
be measured by the {\em relative pulsation amplitude}
defined as 
\begin{equation}
P = \frac{2(C_2^2-C_3^2)}{N} \,.
\label{Pana}
\end{equation}
This quantity varies from $P=0$ for no energy exchange to $P=1$
for maximal exchange.  For $H=0$, it reduces to $P=1-J/N$.
Given the invariants $N$, $H$ and $J$, we may compute $P$
by solving the cubic equation $\Phi(Z)=0$ where, as before,
$\Phi(Z)=\Phi_0(Z)-H^2$, with $Z=|C|^2$ and
$\Phi_0$ defined by (\ref{defPhi0}).
For determination of the envelope, (\ref{Pana}) is ideal.
However, for the inverse problem, it must be simplified.
Noting that $ C_1^2+C_2^2+C_3^2=N$, we may write the pulsation
amplitude as
\[
P = \frac{2(2C_2^2+C_1^2-N)}{N}  \,.
\]
We have already introduced in (\ref{Quadpsi}) a quadratic $\Psi_0$
which approximates the cubic $\Phi_0$ in the range $[C_2^2,C_1^2]$.
If we use the roots of $\Psi_0-H^2=0$ as estimates of $C_1^2$ and $C_2^2$, 
an approximate expression for $P$ may be obtained:
\begin{equation}
\fbox{$
P = 1 -  \sqrt{ \displaystyle{\frac{J^2}{N^2}+\frac{4H^2}{N^2Z_0}}}  \,. 
$} 
\label{Papprox}
\end{equation}
For fixed $P$ this represents an ellipse in $(J,H)$-space. 
The great advantage of (\ref{Papprox}) is that it can be
used to solve the inverse problem.
Two special cases follow immediately:
when $H=0$ then $P=1-J/N$ (which is exact); when $J=0$
then $P=1-2H/N\sqrt{Z_0}$ (which is not exact).

%
\Figure{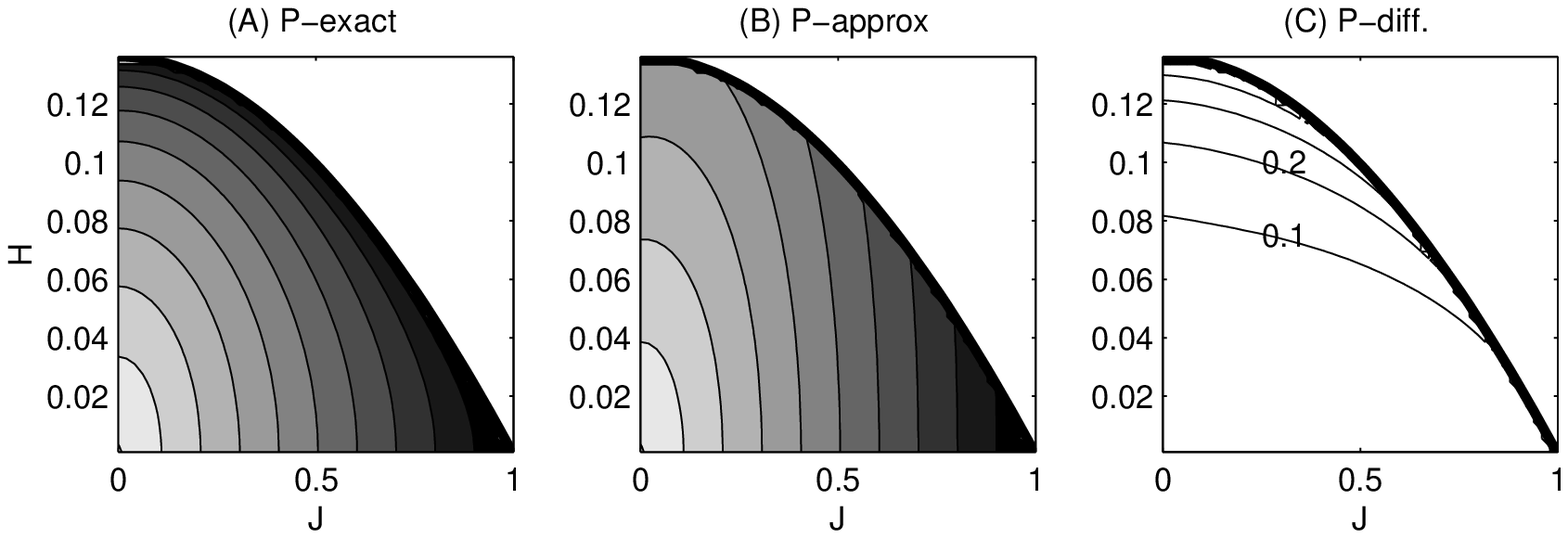}{0.80}{Fig:Pamp}
{Pulsation amplitude.
(A) $P$ based on solving the cubic equation and using (\protect\ref{Pana}).
(B) $P$ from approximation (\protect\ref{Papprox}).
(C) Magnitude of difference between exact and approximate values
The heavy curve is $H_0(J)$.
The contour interval is $ 0.1^\circ$ in all panels.}
%

We plot the exact values of the pulsation amplitude, obtained by
solving the cubic equation $\Phi(Z)=0$, in Fig.~\ref{Fig:Pamp}(A).
Note that $P=0$ when $H=H_0$ and $P=1$ when $H=J=0$.
The approximate values calculated using (\ref{Papprox}) are
shown in Fig.~\ref{Fig:Pamp}(B) and the difference
$(P_{\rm approx}-P_{\rm exact})$ in Fig.~\ref{Fig:Pamp}(C).
The approximation is quite accurate when
$P$ is large. This is the region of primary physical
interest, corresponding to strongly pulsating motion.
For large values of  $H$, the approximation is no longer valid.
We have derived several other approximate expressions for $P$,
which are more accurate, but also more complicated,
than (\ref{Papprox}).

\subsection{Control of the envelope dynamics}


\Figure{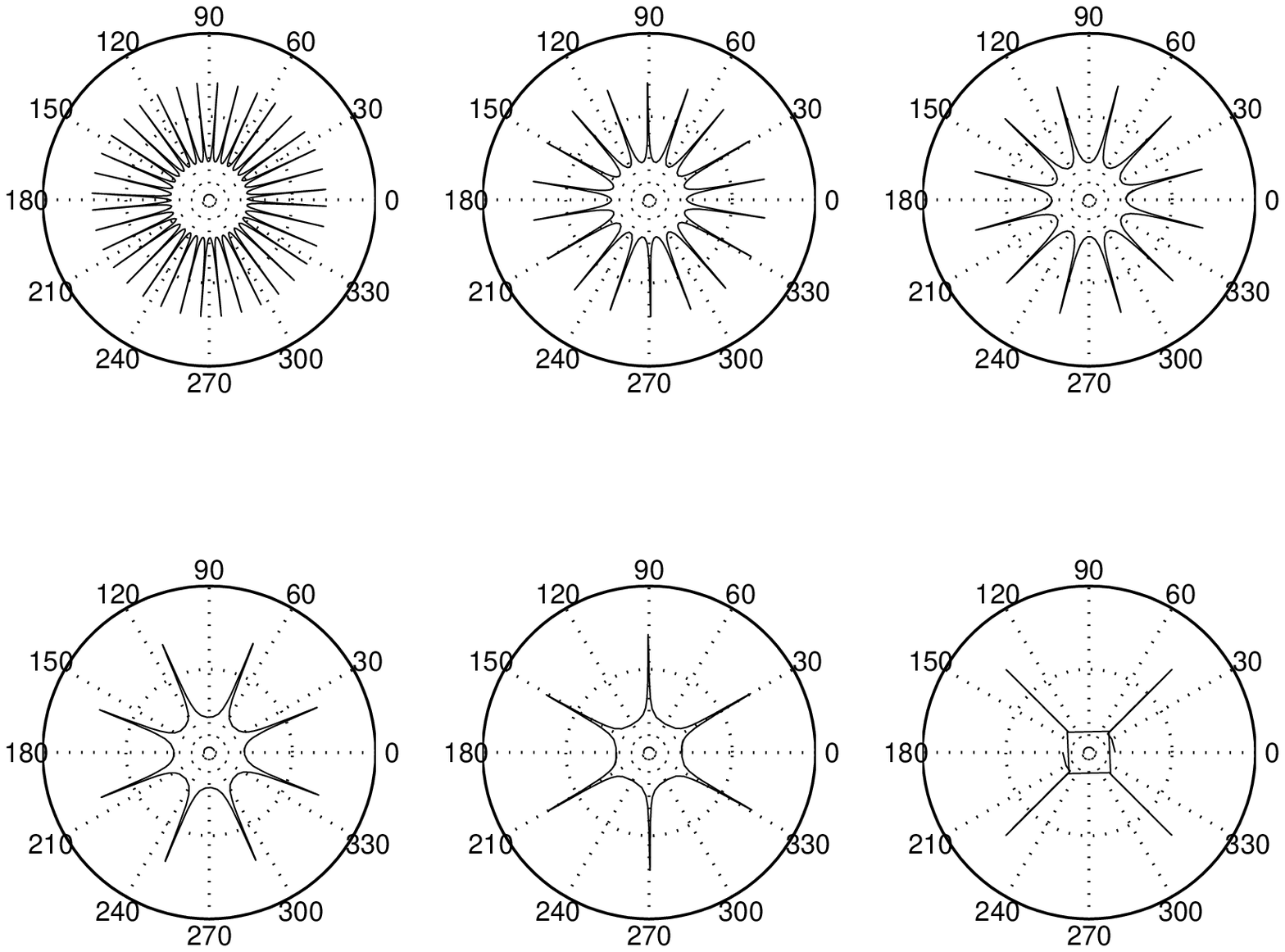}{0.8}{Fig:Blooms}
{Polar plots of $A_{\rm maj}=|A|+|B|$ against $\phi$, computed from the
numerical solution of (\protect\ref{TWEa})--(\protect\ref{TWEc})
for six sets of initial conditions.
For all cases, $N=1$ and $P=0.9$, and $J$ and $H$ are computed from
(\protect\ref{InverseJH}).
Top panels: $\Delta\phi\in\big\{10^\circ, 20^\circ, 30^\circ\big\}$,
bottom panels: $\Delta\phi\in\big\{45^\circ, 60^\circ, 90^\circ\big\}$.
The integration
time in each case corresponds to a total precession of approximately
$180^\circ$, and both $A_{\rm maj}$ and $-A_{\rm maj}$ are plotted.}


The approximate formulas allow us to control the pulsation and
precession by a judicious choice of initial conditions. Recall that the
precession angle is given, to high accuracy, by (\ref{Optform}), which we
write
\begin{equation}
\tan\Delta\phi =          
\frac{J\sqrt{Z_0}}{2H}    
\label{OptformZ0}
\end{equation}
where $Z_0=(N+\alpha J)/2$. 
This may be used in (\ref{Papprox}) to eliminate either $H$ or $J$,
yielding the two equations
\[
P = 1-\frac{J}{N}\csc\Delta\phi 
\qquad \hbox{\rm and} \qquad
P = 1-\frac{2H}{N\sqrt{Z_0}}\sec\Delta\phi \,.
\]
But these are instantly invertible, to give equations for $J$ and $H$
in terms of $P$ and $\Delta\phi$:
\begin{equation}
\fbox{$
J = N(1-P)\sin\Delta\phi
\qquad\hbox{\rm and}\qquad
H = \frac{\sqrt{Z_0}}{2}N(1-P)\cos\Delta\phi \,.
$}
\label{InverseJH}
\end{equation}

To illustrate the effectiveness of these formulas,
six values of the precession angle were
chosen: $\Delta\phi\in\big\{10^\circ, 20^\circ, 30^\circ,
                            45^\circ, 60^\circ, 90^\circ\big\}$.
We set $N=1$ and  fixed the value of the pulsation amplitude
to be $P=0.9$. We then calculated $J$ and $H$ from (\ref{InverseJH})
and computed the
numerical solution of the three-wave equations (\ref{TWEa})--(\ref{TWEc}).
The initial value of $|C|^2$  was taken to be the root $C_2^2$ of
$\Phi(|C|^2)=0$
having intermediate algebraic magnitude. Then $|A|^2$ and $|B|^2$
were obtained from the Manley-Rowe relations. The initial phases were
all set to zero.  Polar plots of $A_{\rm maj}$
against $\phi$ are shown in Fig.~\ref{Fig:Blooms} (the integration
time in each case corresponds to a total precession of about
$180^\circ$, and both $A_{\rm maj}$ and $-A_{\rm maj}$ are plotted).
These plots represent the outer envelope of the horizontal projection
of the trajectory of the pendulum bob.
It is clear that the precession for the numerical solution is,
in each case, close to the value used in
(\ref{InverseJH}).
We also calculated the pulsation amplitude of the
numerical solution and it was, in all cases, within 2\% of the
prescribed value $P=0.9$.
This confirms the effectiveness of the inversion formulas as
a means of pre-determining the envelope of the motion. 

We note that, in general, the horizontal projection of the trajectory
is not a closed curve, but densely covers a region of phase-space.
The motion is not periodic but quasi-periodic. The horizontal
projection is a closed curve only in the exceptional cases when
$\Delta\phi$ and $2\pi$ are commensurate, that is, when their ratio is
a rational number. In this case the motion is periodic and
the horizontal footprint is a star-like graph, as illustrated in
Fig.~\ref{Fig:Blooms}. The number of points in the star is the
denominator of the rational number $\Delta\phi/2\pi$ ({\it e.g.},
$\Delta\phi=80^\circ$ yields a nine-pointed star).


\section{Conclusion}
\label{Sec:Conclusion}

%
\Figure{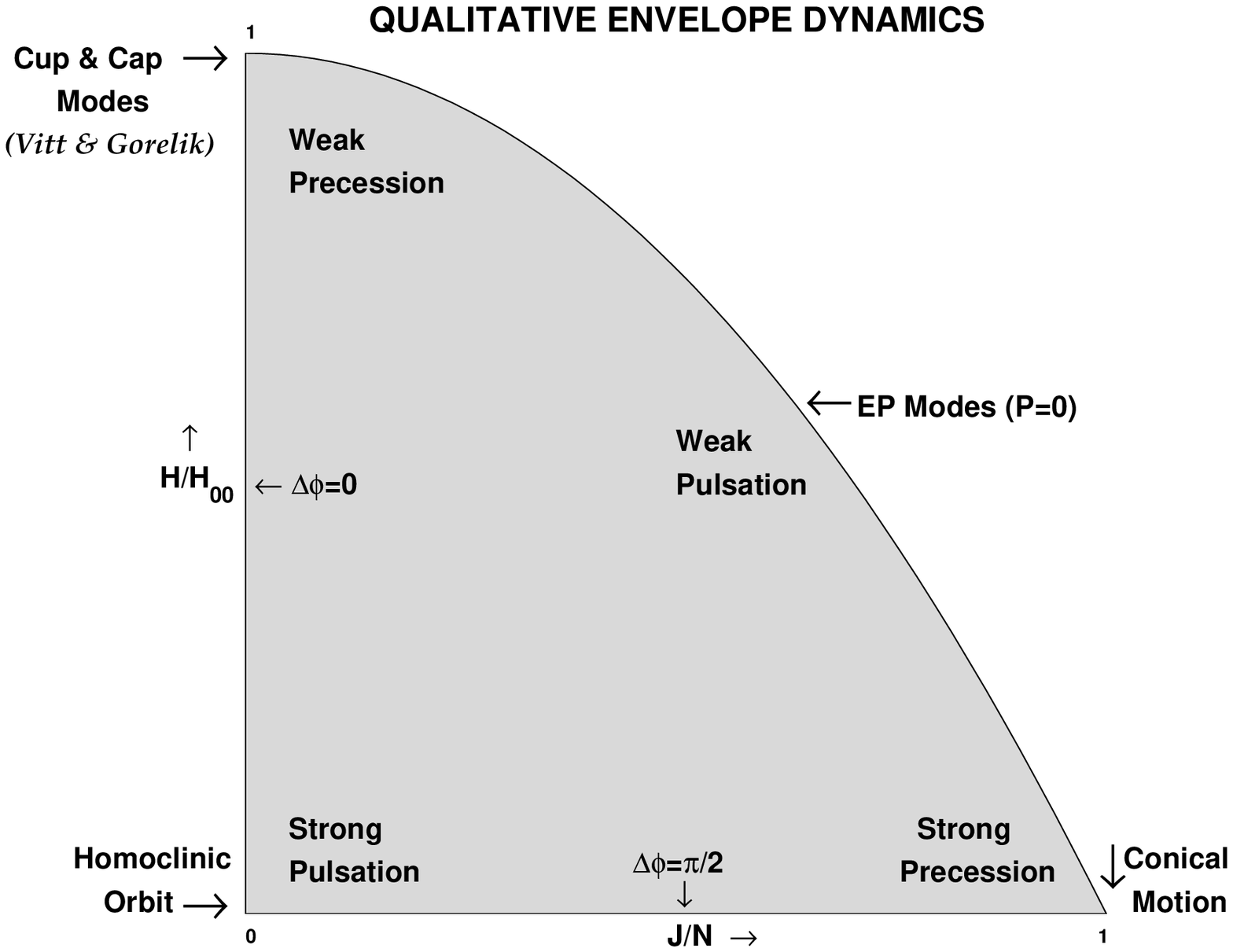}{0.70}{Fig:QED}
{Qualitative features of the envelope dynamics
of the swinging spring.}

We have presented a complete analytical solution of the three-wave
equations, which govern the small-amplitude dynamics of the resonant
swinging spring.  The periodic variation of the amplitudes is associated
with the characteristic pulsation and precession of the system.
Several analytical formulas for the precession angle have been presented.
We have also derived simplified approximate expressions in terms
of elementary functions. The optimal approximation (\ref{Optform})
has been shown by numerical experiments to be remarkably accurate,
with a maximum error of only $0.4^\circ$.
The amplitude of the pulsation envelope is determined from 
the roots of a cubic equation whose coefficients are defined by
the invariants.
Thus, we have provided a complete and positive answer to
Question~1 posed in the Introduction.

The inverse question, Question~2 in \S\ref{Sec:Introduction},
has also been answered affirmatively. 
The approximate formulas (\ref{InverseJH})
give values of $J$ and $H$ which lead to a solution having 
the prescribed pulsation amplitude and precession angle. 
They are of high accuracy for strongly pulsating motion,
which is the case of primary physical interest. 

The qualitative features of the envelope dynamics of the swinging
spring are depicted schematically in Fig.~\ref{Fig:QED}.
The axes are normalized angular momentum $J/N$ 
and normalized Hamiltonian $H/H_{00}$.
The physically accessible domain is shaded. The bounding curve
is $H=H_0(J,N)$. The pulsation amplitude vanishes on this curve
and the solutions are the elliptic-parabolic modes \cite{Lynch2002a}.
Regions of the parameter space are indicated where the pulsation amplitude
and precession angle take large or small values.
The corners of the accessible region represent special solutions.
Thus, $(J,H)=(0,H_{00})$ corresponds to the cup-like and cap-like
solutions of Vitt and Gorelik \cite{V&G}.
For $(J,H)=(N,0)$, the motion of the spring traces out a cone. 
Finally $(J,H)=(0,0)$ represents the homoclinic orbit, and includes 
the case of (unstable) pure vertical oscillations.


\section*{Acknowledgements}

The authors are grateful to Holger Dullin, Loughborough 
University, for informing them of the results derived
by Dullin, Giacobbe and Cushman \cite{Dullin}, and for
providing a preprint of their work.


\appendix


\section{The Nahm Equations and the Three-wave Equations}
\label{Sec:Appendix-A}

The Nahm equations are a set of integrable equations for a three-vector
of skew-Hermetian $n\times n$ matrices $(T_1(s),T_2(s),T_3(s))$:
\begin{equation}
\frac{d}{ds}T_i=[T_j,T_k] = T_j T_k - T_k T_j
\end{equation}
where $(i\,j\,k)$ is a cyclic permutation of $(1\,2\,3)$. In the
simplest case $s\in (-1,1)$ and the matrices have simple poles as
$s\rightarrow \pm 1$.

The Nahm equations were originally discovered because it is possible
to use solutions to the equations to construct solutions of the
Bogomolny equation  \cite{Nahm}. These solutions are called
Bogomolny-Prasad-Sommerfield monopoles. The Bogomolny
equation occurs as a super-symmetry or minimum energy
condition in Yang-Mills Higgs theory and is of interest to theoretical
particle physicists.

There is a Lax formulation of the Nahm equations and an associated Lax
curve of genus $(n-1)^2$.      
The $n=2$ case is elliptic
and the solutions are elliptic functions; in fact, for $n=2$ the Nahm
equations reduce to the Euler-Poinsot equations and are easily
solved. Surprisingly, it is sometimes also the case that the
Nahm equations for $n>2$ can be solved in terms of elliptic functions.
This happens
when the solution has a symmetry and the quotient of the Lax curve by
that symmetry gives a genus-one surface. These symmetries of the Nahm
matrices correspond to spatial symmetries of the corresponding
monopoles. The group elements act both by conjugation on the Nahm
matrices and by rotation of the three-vector of matrices \cite{HMM,HS}.

One example is $n=3$ $D_2$ symmetry \cite{Houghton}.
The symmetry reduces the Nahm matrices to
\begin{eqnarray}
T_1&=&\frac{i}{\sqrt{2}}\left(\begin{array}{ccc} 0 &F_1^* & 0\\
    F_1 & 0 & F_1\\ 0&
    F_1^* & 0 \end{array} \right),\qquad
T_2=\frac{1}{\sqrt{2}}\left(\begin{array}{ccc} 0 & F_2 & 0\\
    -F_2^* & 0 & F_2^*\\ 0&
    -F_2 & 0 \end{array} \right),\nonumber\\[15pt]
T_3&=&\left(\begin{array}{ccc}-i\Re\{F_3\} & 0 &
    -\Im\{F_3\}\\ 0 & 0 & 0 \\ \Im\{F_3\} &0 &
    i\Re\{F_3\}\end{array}\right),\nonumber
\end{eqnarray}
 Substituting these matrices into the Nahm equations gives
\begin{equation} \frac{dF_1}{ds}=F_2^*F_3^*
\label{complexeuler}
\end{equation}
and two others by cyclic permutation. These equations are the
\lq explosive interaction' \hbox{three-wave} equations identified
in \cite{Alber}.  They are related to
the equations studied in the present paper by
$s = it$ with $F_1=A^*$, $F_2=B^*$ and $F_3=C$.


\section{Relationship between Jacobi and Weierstrass
forms of the Precession Angle.}
\label{Sec:Appendix-B}

To relate the expression (\ref{dphi}) obtained by means of Jacobi's
elliptic functions to the expression (\ref{delphi1}) in terms of the
Weierstrass form, we introduce auxiliary
constants $d_{+}$ and $d_{-}$ defined by 
\[
\sn^2d_{+} = \gamma_{+}^2/k^2 \,, \qquad
\sn^2d_{-} = \gamma_{-}^2/k^2 \,.
\]
Note that since $\gamma_{\pm}^2>k^2$, these constants are complex
($d_{+}$ and $d_{-}$ lie on the line between $K$ and $K+iK^\prime$).
It follows that
\[
\sn^2 d_{+} = \left(\frac{e_1-e_3}{e_{+}-e_3}\right)   \,, \quad
\cn^2 d_{+} =-\left(\frac{e_1-e_{+}}{e_{+}-e_3}\right) \,, \quad
\dn^2 d_{+} = \left(\frac{e_{+}-e_2}{e_{+}-e_3}\right) \,,
\]
with similar expressions involving $d_{-}$.
The first of (\ref{xieta2}) may be written
\[
\frac{d\xi}{ds} = -\lambda_{+} 
- \frac{\lambda_{+}\gamma_{+}^2\,\,\sn^2 s }{1-\gamma_{+}^2 \,\sn^2 s}
\,,  \qquad
\]
It may be shown without difficulty, using equation (\ref{Weier2}), that
\[
\lambda_{+}\gamma_{+}^2 = +ik^2\,\sn\,d_{+}\,\cn\,d_{+}\,\dn\,d_{+} \,.
\]
Then the solution (\ref{xixi01}) for $\xi$ may be written
\begin{equation}
\xi - \xi_0 = - \lambda_{+} s - i \Pi_1(s,d_{+},k) \,.
\label{xixi02}
\end{equation}
where $\Pi_1(s,d_{+},k)$ is another standard form (Jacobi's form)
for the elliptic integral
of the third kind (\cite{WW}, \S22.74):
\[
\Pi_1(s,d_{+},k) = 
\int_0^s \frac{ k^2\,\sn\,d_{+}\,\cn\,d_{+}\,\dn\,d_{+}\,\sn^2 s}
                 {1- k^2\,\sn^2d_{+}\,\sn^2 s}  ds \,.
\]

The elliptic integral of the second kind is defined 
(with $x=\sn\,z$) as
\[
E(z) \equiv \int_0^z \dn^2 z \, dz =
\int_0^x \sqrt{\frac{1-k^2x^2}{1-x^2}} \, dx \,.
\]
The complete integral is denoted $E=E(K)$.
$E(z)$ is not periodic; the periodic component is represented by
Jacobi's zeta function
\[
Z(z) = E(z) - E z / K \,.
\]
This is an odd function with period $2K$.
It is related to Jacobi's theta function, also having period $2K$, by
\[
Z(z) = \frac{d}{dz} \log\Theta(z) \,.
\]
The elliptic integral of the third kind may now be expressed as
follows:
\begin{equation}
\Pi_1(z,a,k) = \frac{1}{2} \log \frac{\Theta(z-a)}{\Theta(z+a)} + z Z(a)
\label{ellipint3}
\end{equation}
For the complete form of the integral, when $z=K$,
the logarithmic term vanishes:
\begin{equation}
\Pi_1(a,k) = K Z(a) =
K\frac{\Theta^\prime(a)}{\Theta(a)} \,,
\label{Cellipint3}
\end{equation}
Using this in (\ref{xixi02}), we obtain the change over a half-period $K$: 
\[
\half\Delta\xi = -K\lambda_{+} - iK Z(d_{+}) \,.
\]
Finally, using the analogous expression for $\Delta\eta$,
we get the precession angle
\begin{equation}
\fbox{$
\Delta\phi =
 -K(\lambda_{+}-\lambda_{-}) - iK\big(Z(d_{+})-Z(d_{-})\big) 
$} 
\label{delphi3}
\end{equation}
This is the change over the period $2K$ (for $s$) or $2K/\nu\sqrt{N}$
(for $t$). The structural similarity between this expression and
the result (\ref{delphi1}) in terms of Weierstrass functions
is immediate.

When $J=0$, we have $\lambda_{+}=\lambda_{-}$ and
$\gamma_{+}=\gamma_{-}$, so that $d_{+}=d_{-}$ and
(\ref{delphi3}) implies $\Delta\phi=0$. 
For $H=0$ we have $\lambda_{+}=\lambda_{-}=0$,
$d_{+}=K$ and $d_{-}=K+iK^\prime$. Then using the relation
\[
Z(u+iK^\prime) = Z(u) - \frac{i\pi}{2K} + \frac{\cn\,u\,\dn\,u}{\sn\,u}
\]
with $u=K$, it follows immediately that $\Delta\phi=\pi/2$, in agreement
with (\ref{dphihalfpi}).


\section{Approximation Integral with
Best Fit at $\mathbf{Z=N_{+}}$.}
\label{Sec:Appendix-C}

We approximate $\Phi_0=Z(Z-N_{+})(Z-N_{-})$
by a quadratic with a root at $Z-N_{+}$:
\begin{equation}
\Psi_0= Z_0(Z-N_{+})(Z - Z_1)
\label{GenPsiAgain}
\end{equation}
To obtain the best fit at $Z=N_{+}$, we choose
$Z_0$ and $Z_1$ so that 
\begin{equation}
\Psi_0(Z)-\Phi_0(Z) = O\left((Z-N_{+})^3\right) \,.
\end{equation}
This implies $Z_0=N_{+}+J$ and $Z_1=N_{+}^2/(N_{+}+J)$.
Using the software package {\sc Maple},
it is possible to evaluate the resulting approximate integral
\begin{equation}
\Delta\phi=\frac{HJ}{2}\int^{\infty}_{C_{+}}
\frac{d|C|^2}{|A|^2|B|^2\sqrt{\Psi_0-H^2}}.
\end{equation}
where $C_{+}$ is the larger root of $\Psi_0-H^2=0$. The result is
\begin{eqnarray}
\Delta\phi&=&\frac{1}{4}
\left(2\tan^{-1}{\frac{J(N+J)}{2\sqrt{2}
\sqrt{N+3J}H}}+\pi\right)\cr
 &&-\frac{H}{4\sqrt{J^3-H^2}}\left(2\tanh^{-1}\,
\frac{J(5J+N)}{2\sqrt{2}\sqrt{J^3-H^2}\sqrt{N+3J}}+i\pi\right)
\end{eqnarray}
This gives a rather good approximation: the maximum error is
about $1.2^\circ$.  However, it is not a convenient expression because 
the factor $\sqrt{J^3-H^2}$ is sometimes real and
sometimes imaginary. Moreover, the expression cannot easily be inverted
to give $H$ or $J$ in terms of $\Delta\phi$.
In fact, unless $Z_1=N_{-}$, any approximating
quadratic (\ref{GenPsiAgain}) will give an expression with this problem.
Thus, in \S5, we choose $Z_1=N_{-}$.




\begin{thebibliography}{00}

\bibitem{A&S}
M. Abramowitz and I.A. Stegun, Handbook of Mathematical Functions,
Dover Publ., New York. 1965.

\bibitem{Alber}
M.S. Alber, G.G. Luther, J.E. Marsden and J.M. Robbins,
Geometric phases, reduction and  Lie-Poisson structure for the resonant
three-wave interaction,
{\it Physica D} {\bf 123} (1998), pp. 271--290.
\hfill\break
{\tt
http://www.cds.caltech.edu/$\sim$marsden/bib\_src/papers/twi\_latex.pdf}

\bibitem{Audin}
M. Audin, Spinning Tops. A Course on Integrable Systems,
Cambridge Univ.~Press, Cambridge, 1996.

\bibitem{Byrd}
P.F. Byrd, P. F. and M. D. Friedman, Handbook of elliptic integrals for
engineers and scientists, Second edn.
Springer-Verlag, Berlin, 1971.

\bibitem{Dullin}
H. Dullin, A.~Giacobbe and R.~Cushman,
Monodromy in the resonant swing spring, preprint, 2002.
\hfill\break
{\tt nlin.SI/0212048}.

\bibitem{Dwight}
H.B. Dwight, Tables of integrals and other mathematical data,
Macmillan, New York, 1947.

\bibitem{Gradshteyn}
I.S. Gradshteyn and I.M. Ryzhik, Table of Integrals, Series and Products,
Acad.~Press, New York. 1965.


\bibitem{HMM} N.J. Hitchin, N.S. Manton and M.K. Murray, Symmetric
monopoles, Nonlinearity {\bf 8} (1995), p. 661. {\tt dg-ga/9503016}.


\bibitem{H&L}
D.D. Holm and P. Lynch, Stepwise precession of the resonant swinging
spring.  SIAM J. Appl. Dynam. Sys. {\bf 1} (2002), pp. 44--64.
\hfill\break
{\tt nlin.CD/0104038}

\bibitem{Houghton} C.J. Houghton, Multimonopoles, (Thesis, Cambridge, 1994).



\bibitem{HS}
C.J. Houghton and P.M. Sutcliffe, Tetrahedral and cubic monopoles, Commun.
Math. Phys. {\bf 180} (1996), p. 342 {\tt hep-th/9601146}.

\bibitem{Lawden}
D.F. Lawden, Elliptic functions and applications,
Springer-Verlag, New York, 1989.

\bibitem{Lynch2002a} P. Lynch, Resonant motions of the
three-dimensional elastic pendulum.  Intl.~J.~Nonlin.~Mech. {\bf 37}
(2002), pp. 345--367.
\hfill\break {\tt
http://www.maths.tcd.ie/$\sim$plynch/Publications/IJNM\_Paper.pdf}

\bibitem{Lynch2002b}
P. Lynch,  The swinging spring: a simple model for atmospheric balance, in
J.~Norbury and I.~Roulstone (Eds.), Large-scale atmosphere-ocean dynamics:
Vol II: geometric methods and models,
 Cambridge University Press, Cambridge, 2002, pp. 64--108.
\hfill\break
{\tt http://www.maths.tcd.ie/$\sim$plynch/Publications/AOD\_Paper.pdf}

\bibitem{Lynch2003}
P. Lynch, Resonant Rossby wave triads and the swinging spring.
To appear in Bull.~Amer.~Met.~Soc. {\bf 84} (2003).
\hfill\break
{\tt http://www.maths.tcd.ie/$\sim$plynch/Publications/RRTSS.pdf}

\bibitem{Nahm} W. Nahm, The construction of all selfdual
multi-monopoles by the ADHM method, in N.S. Craigie, P. Goddard and
W. Nahm, (Eds.), Monopoles in quantum field theory, World Scientific,
Singapore, 1982, p. 87.

\bibitem{V&G}
A. Vitt and G. Gorelik,
Kolebaniya uprugogo mayatnika kak primer kolebaniy
dvukh parametricheski svyazannykh linejnykh sistem.
 Zh.~Tekh.~Fiz. (J.~Tech.~Phys.) {\bf 3}(2-3) (1933), pp. 294--307.
Available in English translation: Oscillations of an elastic pendulum as
an example of the oscillations of two parametrically coupled linear
systems,
translated by Lisa Shields, with an introduction by Peter Lynch,
Historical Note No.~3, Met \'Eireann, Dublin (1999).
\hfill\break
{\tt http://www.maths.tcd.ie/$\sim$plynch/Publications/VandG.pdf}

\bibitem{Whittaker37}
E.T. Whittaker,
A Treatise on the Analytical Dynamics of Particles and
Rigid Bodies. 4th Edn. Cambridge Univ.~Press, Cambridge 1937.

\bibitem{WW}
E.T. Whittaker and  G. N. Watson, A Course of Modern Analysis.
4th Edn.  Cambridge Univ.~Press, Cambridge 1927.


\end{thebibliography}
\end{document}